# Suhl Instabilities in Nanoscopic Spheroids

Jinho Lim[1], Anupam Garg[1], and John B. Ketterson[1,2]

1. Department of Physics and Astronomy, Northwestern University, Evanston, IL, 60208.
2. Department of Electrical and Computer Engineering Northwestern University, Evanston, IL, 60208.

## Abstract

We simulate the magnetization dynamics of a permalloy spheroid of nanoscopic size in zero external field, such that both dipolar and exchange interactions are important. Low excitation power is used to obtain the frequencies and mode patterns of many normal modes. At higher power, non-linear three and four mode couplings between magnons carrying orbital angular momentum are observed to give rise to Suhl instabilities. Suhl's analysis of the selection rules governing the allowed processes is extended to initial states other than uniform precession. These rules are studied and confirmed by the simulations. Both down- and up-conversion are seen as well as three and four-mode processes. General trends are inferred for preferred instabilities among those that are allowed, although the thresholds for some instabilities appear to be very high.

## I. Introduction

Early ferromagnetic resonance (FMR) experiments at large amplitudes performed by Damon [1] and by Bloembergen and Wang [2] displayed two then-unexpected effects: 1) a premature saturation of the primary absorption at an applied power that was much smaller than that expected on the basis of the line width of the resonance, and 2) a second (auxiliary), relatively broad, absorption feature [3]. The origin of these two effects was later studied in some detail by H. Suhl [4, 5] and shown to arise from a non-linear coupling between the uniformly precessing magnetization and certain nonuniform magnetization modes (spin waves) of the system. The auxiliary feature was shown to arise from a parametric coupling between various modes at the applied frequency and two spin waves having half that frequency, while the premature saturation arose from a coupling between the resonant uniform precession and two spin waves having the same frequency.

Suhl's analysis was carried out under the assumption that the nonuniform modes are plane-wave like and numbered by their wave vectors, or lattice momentum, **k**. Conservation of this quantity then gives the following selection rules:

$$\omega(\mathbf{k}) + \omega(-\mathbf{k}) = \omega_0, \qquad \text{(Process 2)}, \qquad (1.1)$$

and

$$\omega(\mathbf{k}) + \omega(-\mathbf{k}) = 2\omega_0, \qquad \text{(Process 1)}. \qquad (1.2)$$

where $\omega_0$ is the initial (applied) frequency and $\omega(\mathbf{k})$ is a resultant frequency. These equations have a simple quantum mechanical interpretation in terms of emission and absorption of magnons and they correspond to three and four mode (magnon) processes respectively. In a purely classical framework, they are instances of the general Manley-Rowe relations [6-8].

In the presence of dipolar and exchange effects there are modes with frequencies, $\omega(\mathbf{k})$, that are the same as $\omega_{FMR}$, the resonance frequency of the uniform precession mode. If $\omega_0 = \omega_{FMR}$, process 1 can occur resonantly, which results in the premature saturation of the FMR resonance noted above. If this condition is not satisfied the strength of this process is greatly reduced.

Regarding process 2, it is the analog of parametric down conversion in nonlinear optics. If $\omega(\mathbf{k})$ possesses an extremum where $d\omega(\mathbf{k})/d\mathbf{k} = 0$, the transitions will be especially strong due to the large density of states leading to an auxiliary absorption feature occurring at a frequency $\omega_0$ (which differs from $\omega_{FMR}$ in general). This latter effect has recently been studied systematically in a thin film over a wide range of frequencies [9]. For special values of $\omega_0$ and H (the applied field) this latter process can occur for $\omega_0 = \omega_{FMR}$ which also contributes to a premature saturation. We note that the presence of frequencies below the uniform mode, as required for Processes 1 and 2 to occur, arises directly from demagnetization effects.

Following Suhl's treatment of the instabilities described by (1.1) and (1.2), there have been various studies of these effects and related nonlinear phenomena. Most previous studies involved films with macroscopic or mesoscopic thicknesses or macroscopic three-dimensional confined geometries (e.g., spheres). Since these sample dimensions were generally much larger than the wavelength of typical spin waves investigated, the language of plane waves is appropriate. Experimental topics investigated include: 1) the Suhl instability threshold (critical driving field amplitude) versus applied static field strength at a fixed driving frequency (the butterfly curve) [10-19], and 2) other nonlinear phenomena above the threshold of the first order Suhl instability (1.1) including various nonlinear FMR responses (fold-over FMR, asymmetric FMR) [20-23] and typical topics in nonlinear dynamics (period doubling, auto oscillation, bifurcation, chaos etc.) [24-29]. There are also various theoretical descriptions of the Suhl instability using 1) the Landau-Lifshitz equation [30-35], 2) classical Hamiltonians [36-42], and 3) quantum mechanical Hamiltonians [43-45]. As already mentioned above, a common feature of these theoretical descriptions is that they are formulated in terms of plane waves.

Central to the use of plane waves is the assumption that the sample is sufficiently large (in the lateral directions for a thin film). The reason is that plane waves are a good approximation for the actual modes only for wavelengths that are small compared to sample dimensions due to the additional complexity introduced by the presence of a demagnetization field, which must be such as to satisfy the Maxwell boundary conditions on **B** and **H** at the sample surface. Additional boundary conditions arising from exchange effects, whose importance grows as the sample gets smaller, must also be obeyed. For slabs of infinite extent, the mode spectrum was calculated by Damon and Eshbach [46] in the absence of exchange effects; the theory was subsequently extended to include the exchange effect [47-50].

With recent developments in 3d nanomagnetism [51-57], sub-micron scale system can be fabricated with a variety of techniques: 1) physical vapor deposition (PVD) onto previously patterned 3D scaffolds [52, 55], 2) chemical synthesis [51, 57], and 3) 3D nano-printing using focused electron beam induced deposition (FEBID) [53]. The study of spin dynamics in such small systems is therefore very timely. Furthermore, physical laws associated with multi-magnon processes are fundamentally related to geometric symmetries and since systems in 3d can have more complex geometries than those in 2d, we expect that there will many fruitful, symmetry-related laws for multi-magnon processes in 3D systems with different symmetries. Here we show such laws for 3, and 4-magnon processes in a spheroid.

For the case of a spheroid, which is perhaps the simplest 3d geometry, analytic calculations of the mode frequencies were carried out by Walker, but in the absence of exchange effects [58]. Extending Walker's analysis of spheroids to include the effects of exchange would be challenging and to our knowledge has not been attempted; even more challenging would be to describe the onset of non-linear effects. To study multi-mode (or multi-magnon) processes or Suhl instabilities in such samples, one needs to use the language of magnetic normal modes rather than plane waves. Therefore, the wavevector conservation rule implicit in eq. (1.1) and (1.2) will be replaced by selection rules associated with the mode numbers and symmetries of the normal modes. Motivated by the observation of a rather nonmonotonic dependence of the Kittel mode frequencey on the applied magnetic field in elliptical disk arrays [59], we successfully simulated resonant and nonresonant 2nd and 3rd harmonic generation in a single elliptical disk, corresponding to processes in which two or three magnons of a low frequency nonuniform cap mode transform into one magnon of the higher frequency Kittel mode [60]. A study (of which we learned well after this research was completed) of three-mode processes has recently been carried for a circular permalloy nanodot of diameter ~5 μm, and thickness 50 nm [61]. The two-dimensional geometry and the high symmetry of the system facilitate calculation of the modes and the selection rules. The exercise is more difficult for a fully three-dimensional particle. In addition, confirming these selection rules experimentally with a real sample would be prohibitive because it is very hard to identify which normal modes are excited in the sample. From this perspective, micromagnetic simulations which can generate high-resolution spin maps are a very attractive way to confirm these selection rules. Also, by choosing a nano-scale spheroid, one can decrease the density of normal modes and this will aid the investigation greatly. Therefore, we will here examine, numerically, Suhl-like processes for some of the low-lying modes of a spheroid (ellipsoid of revolution) primarily with dimensions (principal axes) 50×50×100 nm$^3$. The spheroid geometry is a good starting point for this kind of complex simulational study because the system can support a uniform dipolar field (or demagnetizing field) (this is well known, but for an elementary derivation, see Garg et al [62]) so that the system doesn't have a class of modes which are confined to a restricted area near the surface, so called edge, localized, or cap modes [63-69] which add more complexity to the study. The analysis can then be conducted with bulk modes only. Due to its favorable magnetic properties, we will use material parameters relevant to permalloy (Py): gyromagnetic ratio $\gamma = 2\pi \times 2.80$ GHz/kOe, saturation magnetization $M_0 = 650$ emu/cm$^3$, damping constant $\alpha = 8 \times 10^{-3}$, and exchange constant $A_{ex} = 1.0 \times 10^{-6}$ erg/cm. Most of our calculations are carried out in the absence of an applied magnetic field where precession occurs under the sample's own demagnetization field.

Our computational approach utilizes the open source OOMMF code [70] developed by the National Institute of Standards and Technology (NIST). We employ a grid (cell) size of 2×2×2 nm$^3$ which is chosen to: 1) be smaller than magnetostatic exchange length, $l_{ex} = \sqrt{A_{ex}/2\pi M_0^2}$ ($\approx$ 6 nm) of the chosen material (Py), 2) provide enough spatial resolution to represent spatial profiles of various modes, and 3) make sure the system has a numerically sufficiently uniform demagnetizing field. Modes generated through nonlinear processes are identified via a cell by cell fast Fourier transform (FFT) of the inhomogeneous time-dependent magnetization resulting from some initially excited spatial pattern, as described by Lim et al. [63]. Through an examination of the spatial dependence of the responses of the individual cells at a common frequency we can obtain the associated patterns and quantum numbers of the modes so excited. For an FMR-like experiment this would be accomplished simply by a uniform tipping of the initial magnetization away from the equilibrium direction (lying parallel to the long axis (the axis of revolution) at zero field) and letting the system evolve in time. We will also examine some initial states having a nonuniform distribution that mimics excitation via a mode other than uniform precession.

The plan of the paper is as follows. In Sec. II we examine the normal modes. In Sec. III we discuss generalized selection rules, thresholds of Suhl instabilities, and frequency shifts due to high amplitudes of excitations. In sections IV, V, and VI, we study the instabilities arising from excitations with orbital angular momentum $p = 0, -1,$ and $1$, respectively, and give a comprehensive discussion of the selection rules, up and down conversion, and three and four-mode processes. General conclusions are given in Sec. VII. Some simple theoretical aspects are analyzed in Appendices A and B.

## II. Low-lying modes in the linear regime

Prior to examining various non-linear mode couplings, we will first examine some of the low-lying low-amplitude modes of our spheroid, following the simulation techniques in Lim et al [63]. As explained earlier, these modes must be found numerically. A formulation of the numerical eigenvalue problem using the mathematical language of Fréchet derivatives and Banach spaces is given in Ref. [71]. Here we will adopt the notation introduced by Lim et al. [63] ($p$, $n_z$, $n_r$), wherein $n_z$ is the number of nodes along the spheroid axis ($z$), $n_r$ is the number of radial nodes (excluding those on the axis), and $p$ is the dominant orbital angular momentum along $z$. The total angular momentum is $p + 1$. A more detailed discussion of why we choose to label the modes with just the orbital part, p, instead of the total angular momentum, m, can be found in Lim et al. [63]. Magnons carrying orbital angular momentum (OAM) have recently gotten much attention [72-76] because $p$ can be identified as a topological charge. Thus, the integral of the OAM over the cross-section is protected against magnetic damping [72, 74]. It is speculated that a magnon with non-zero $p$ may carry intrinsic angular momentum of $\sim \hbar p$ [74].

Although normal modes can be labeled by only one index as in Ref. [71], our indexing scheme is central to properly representing the symmetries of the modes and enable the analysis of selection rules associated with the instabilities. The modes have definite parity in $z$, which is the same as that of $n_z$. In most cases we have $n_r = 0$, and then we will omit $n_r$ and use the label ($p$, $n_z$). The modes are excited by evolving an inhomogeneous initial magnetization distribution in the absence of damping ($\alpha$ set to zero) and observing the frequencies and the associated mode patterns of the excitations so generated. The resulting frequencies together with their assigned mode numbers are collected in Table I. Note that both the dipolar- and exchange-interactions contribute here. Since $\omega(0, 0, 0) = 5.66$ GHz $> \omega(0, 1, 0) = 4.98$ GHz, which is typical of backward-volume like behavior [46, 77], and since $p = \pm 1$ modes are significantly nondegenerate [63], (especially for small $n_z$), one can conclude that the dipolar interaction is not negligible in our sample with the given dimensions. The increasing effect of exchange interaction at a fixed $n_r$ and p is seen in Table

I as a monotonic increase in frequency with increasing $n_z$ for $n_z \geq 2$ for $(p, n_r) = (0, 0)$, and all $n_z$ for $(p, n_r) \neq (0, 0)$.

Table I. Low-lying normal mode frequencies

| $p$ | $n_z$ | $n_r$ | frequency (GHz) | $p$ | $n_z$ | $n_r$ | frequency (GHz) |
|---|---|---|---|---|---|---|---|
| 0 | 0 | 0 | 5.66 | 2 | 0 | 0 | 21.48 |
| 0 | 1 | 0 | 4.98 | 2 | 1 | 0 | 25.59 |
| 0 | 2 | 0 | 6.64 | 2 | 2 | 0 | 31.05 |
| 0 | 3 | 0 | 10.06 | 2 | 3 | 0 | 37.79 |
| 0 | 4 | 0 | 15.33 | 2 | 4 | 0 | 45.90 |
| 0 | 5 | 0 | 22.17 | | | | |
| 0 | 6 | 0 | 30.66 | –3 | 0 | 0 | 31.93 |
| 0 | 7 | 0 | 39.55 | –3 | 1 | 0 | 38.18 |
| | | | | –3 | 2 | 0 | 45.21 |
| –1 | 0 | 0 | 6.45 | –3 | 3 | 0 | 53.22 |
| –1 | 1 | 0 | 11.23 | | | | |
| –1 | 2 | 0 | 16.21 | 3 | 0 | 0 | 33.20 |
| –1 | 3 | 0 | 21.88 | 3 | 1 | 0 | 39.06 |
| –1 | 4 | 0 | 28.52 | 3 | 2 | 0 | 46.00 |
| –1 | 5 | 0 | 35.94 | 3 | 3 | 0 | 53.71 |
| –1 | 6 | 0 | 45.61 | | | | |
| | | | | 0 | 0 | 1 | 28.42 |
| 1 | 0 | 0 | 12.30 | 0 | 2 | 1 | 42.19 |
| 1 | 1 | 0 | 14.36 | 0 | 4 | 1 | 61.33 |
| 1 | 2 | 0 | 17.87 | | | | |
| 1 | 3 | 0 | 22.95 | 1 | 0 | 1 | 50.00 |
| 1 | 4 | 0 | 29.10 | | | | |
| 1 | 5 | 0 | 36.33 | | | | |
| 1 | 6 | 0 | 45.90 | | | | |
| | | | | | | | |
| –2 | 0 | 0 | 19.04 | | | | |
| –2 | 1 | 0 | 23.93 | | | | |
| –2 | 2 | 0 | 29.59 | | | | |
| –2 | 3 | 0 | 36.43 | | | | |
| –2 | 4 | 0 | 44.43 | | | | |

## III. Selection rules, thresholds, and frequency shifts

The nonlinear effects we see in spheroidal samples are governed by general selection rules which we derive in Appendix A. Suhl also stated [5] a selection rule, but that was limited to a specific kind of instability occurring from the uniformly excited state, i.e., the Kittel mode and also did not distinguish spin and orbital angular momentum of magnons. The more general rules for a three-mode process are

$$\omega_i = \omega_{f,1} + \omega_{f,2} \quad \text{or} \quad \omega_{i,1} + \omega_{i,2} = \omega_f, \tag{3.1}$$

$$p_i - 1 = p_{f,1} + p_{f,2} \quad \text{or} \quad p_{i,1} + p_{i,2} = p_f - 1, \tag{3.2}$$

and

$$n_{z,1} + n_{z,2} + n_{z,3} = 1 \ (\text{mod } 2). \tag{3.3}$$

Here, the subscripts 1, 2, and 3 refer to the three modes involved, and $i$ and $f$ refer to initial and final states. Eq. (3.2) arises from conservation of total angular momentum and is the counterpart of conservation of **k** (linear momentum) for the bulk (or unbounded system) case. Every magnon has intrinsic spin angular momentum of 1 (in units of $\hbar$) and it is exchanged with OAM during the 3-magnon process through the dipolar interaction even in the absence of spin-orbit coupling [73]. The rule (3.3) is the parity rule. The first and second cases of eq. (3.1) and (3.2) will be called, respectively, three-mode down and up conversion in the sense of parametric down and up conversion in nonlinear optics.

To test these selection rules, we will use three different excitation fields that can excite modes with $p_i = 0, \pm 1$. The corresponding simulations are described in Secs. IV–VI.

The threshold field $H_{1,t}$ for the process $i \rightarrow 1 + 2$ is derived in Appendix B (with the modes labelled as 1, 2, and 3 instead of $i$, 1, and 2), in Eq. (B.7). It depends on two detunings (see Appendix B): $\omega_0 - \omega_i$, and $\omega_0 - \omega_1 - \omega_2$, where $\omega_0$ is the frequency of the applied microwave field, and $\omega_1$ and $\omega_2$ are the natural frequencies of modes 1 and 2, which may differ from their linear regime values, because when the instability occurs these modes are excited on a background of a preexisting mode $i$ of potentially large amplitude. Our general observation is that the mode frequencies are pulled downward, but this is not a universal trend.

Second, $H_{1,t}$ depends on the individual mode damping rates $\eta_j$,

$$\eta_j = \alpha \omega_j K_j, \tag{3.4}$$

where $\alpha$ is the Landau-Lifshitz damping parameters (assuming $\alpha \ll 1$), and $K_j$ is a number of order 1 reflecting an average precessional ellipticity of the mode. By this we mean that in general the vector $\mathbf{m}(\mathbf{r},t)$ traces out an ellipse and not a circle as it precesses, and this ellipse may be less or more eccentric at different $\mathbf{r}$.

The most significant dependence is that $H_{1,t}$ varies inversely with the coupling constant $g_{i,12}$. These coupling constants must be found numerically, but generally we expect them to be smaller if modes with higher quantum numbers (and thus more nodes) are involved. We comment further on this point in Sec. IV B.

Finally, we also find in Appendix B that exactly at threshold, the modes 1 and 2 are excited at frequencies

$$\omega'_{1,2} = \omega_{1,2} + \frac{\eta_{1,2}}{\eta_1 + \eta_2} \zeta, \tag{3.5}$$

where

$$\zeta = \omega_0 - (\omega_1 + \omega_2).$$

The Manley-Rowe condition is still obeyed, but now reads

$$\omega'_1 + \omega'_2 = \omega_0. \tag{3.6}$$

## IV. Three-mode instabilities with a uniform excitation field

To excite a mode like the uniform FMR mode with $p_i = 0$ one needs to use quasi-uniform, time-varying magnetic fields and one doesn't need a special apparatus to make these fields in a lab. For example, a single-wire antenna or a coplanar waveguide can do this job effectively. In the simulations, we apply a circularly polarized, spatially uniform, $H_1$ field in the plane perpendicular to the axis of revolution of the spheroid (the z-axis) and oscillating with a fixed frequency. Specifically, we take

$$\mathbf{H}_{\text{mw}}(t) = H_1(t)(\cos\omega_0 t\ \hat{\mathbf{x}} + \sin\omega_0 t\ \hat{\mathbf{y}}), \tag{4.1}$$

where the amplitude $H_1(t)$ is allowed to be time dependent to enable the initial excitation amplitude to slowly grow. We take

$$H_1(t) = \dot{H}_1 t, \tag{4.2}$$

where $\dot{H}_1$ has a default value of 4 Oe/ns but can be as big as 40 Oe/ns. By monitoring the transverse magnetizations and the exchange energy (which is an excellent measure of the inhomogeneity in **M**) as a function of time, we can roughly detect the time when the instability occurs (or times if there is more than one instability). Fig. 1 shows raw data of a simulation with a microwave field (4.1) corresponding to Case 3 in Table II. As inhomogeneous modes which have different symmetries than the initial mode rapidly grow on top of the initial mode, the system's exchange energy also grows rapidly. Let the time of instability found in this way be $t_i$. We then conduct a cell by cell FFT just as in Sec. II in order to get frequencies and spatial patterns of newly excited modes. The FFT is done over a time window $(t_i - \tau_1, t_i + \tau_2)$, where the offsets $\tau_1$ and $\tau_2$ are chosen so as to capture a significant portion of the post-instability evolution but not so much as to overlap with the period when the development has become chaotic. Typically, $\tau_1 + \tau_2 = 10$ ns and the sampling is done every 10 ps. The corresponding Nyquist frequency, $f_N$, is 50 GHz, and a mode with frequency $f > f_N$ will be aliased to $\bar{f} = \min(\tilde{f}, 2f_N - \tilde{f})$ where $\tilde{f} \equiv f \pmod{2f_N}$ (see, e.g. chapter 12 of [78]). The concepts of Nyquist frequency and aliasing are widely used in physical sciences. For example, in a periodic crystal the first Brillouin zone (FBZ) boundary is a Nyquist frequency of data sampled from lattice sites, and Umklapp scattering is an aliasing because the final wavevector is outside of the FBZ. Fig. 2 shows two power spectral densities (PSD) at different time windows: (a) from 0 to 10 ns and (b) from 30 to 40ns. A definition of $\bar{S}_x(\omega)$ can be found as Eq. (40) in Lim et al. [63]. We see that an initial (0, even, 0) mode at $f_0 = 10.64$ GHz splits into two modes with quantum numbers as shown. These quantum numbers are found by examining the associated patterns shown in Figs. 3(a)–(e). From Fig. 3(b) we see that the (0, even, 0) mode mainly consists of the (0, 0, 0) mode and a less than 1% contamination of the (0, 2, 0) pattern. Note that the x-y cross sections show both x- and y-components of the mode amplitude function as vectors, but the y-z cross sections only show its x-component. A detailed discussion of the definition of the mode amplitude function and meaning of its real and imaginary parts is given near Eq. (42) in Lim et al. [63].

We note that identifying the onset of instability and choosing a good time window for spectral analysis is neither straight forward nor systematic. Once this is done, however, it is generally straight forward to identify resonances and their associated spatial patterns, except in cases where multiple patterns are found at a single frequency as shown in Fig. 3(b).

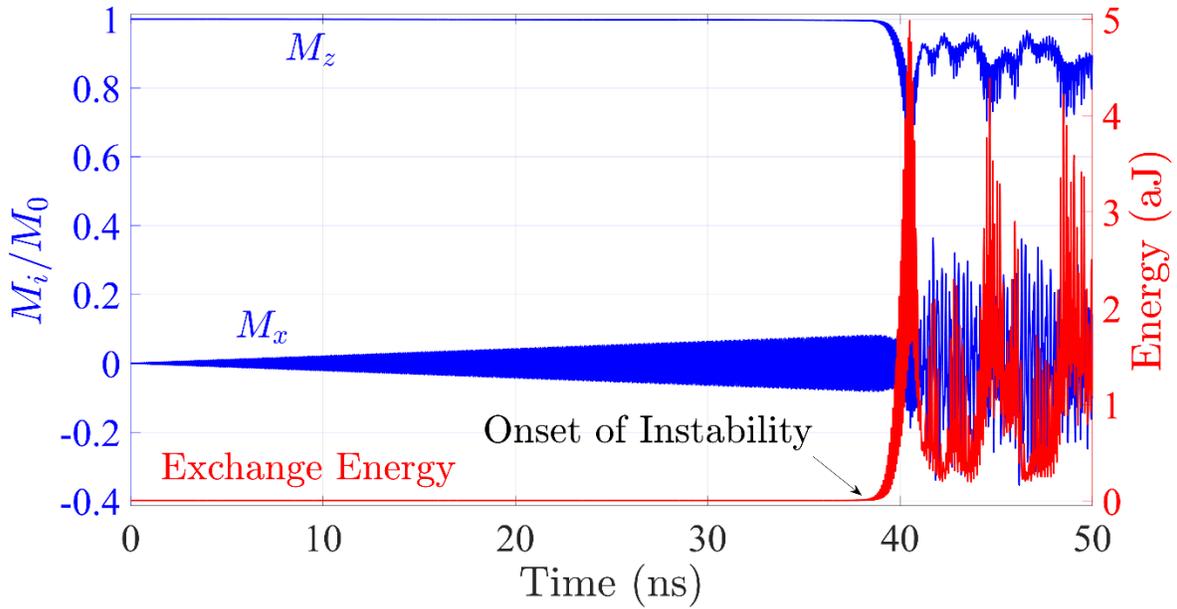

FIG. 1. A simulation with a driving field of the form (4.1) with $\dot{H}_1 = 4$ Oe/ns, and $f_0 = \omega_0/2\pi = 10.64$ GHz. $M_x$ and $M_z$ are $x$- and $z$-components of the magnetization of the whole system respectively, and $M_0$ is the magnitude of the magnetization. The exchange energy starts to grow rapidly around 38 ns which corresponds to onset of the three-mode instability of Case 3 in Table II.

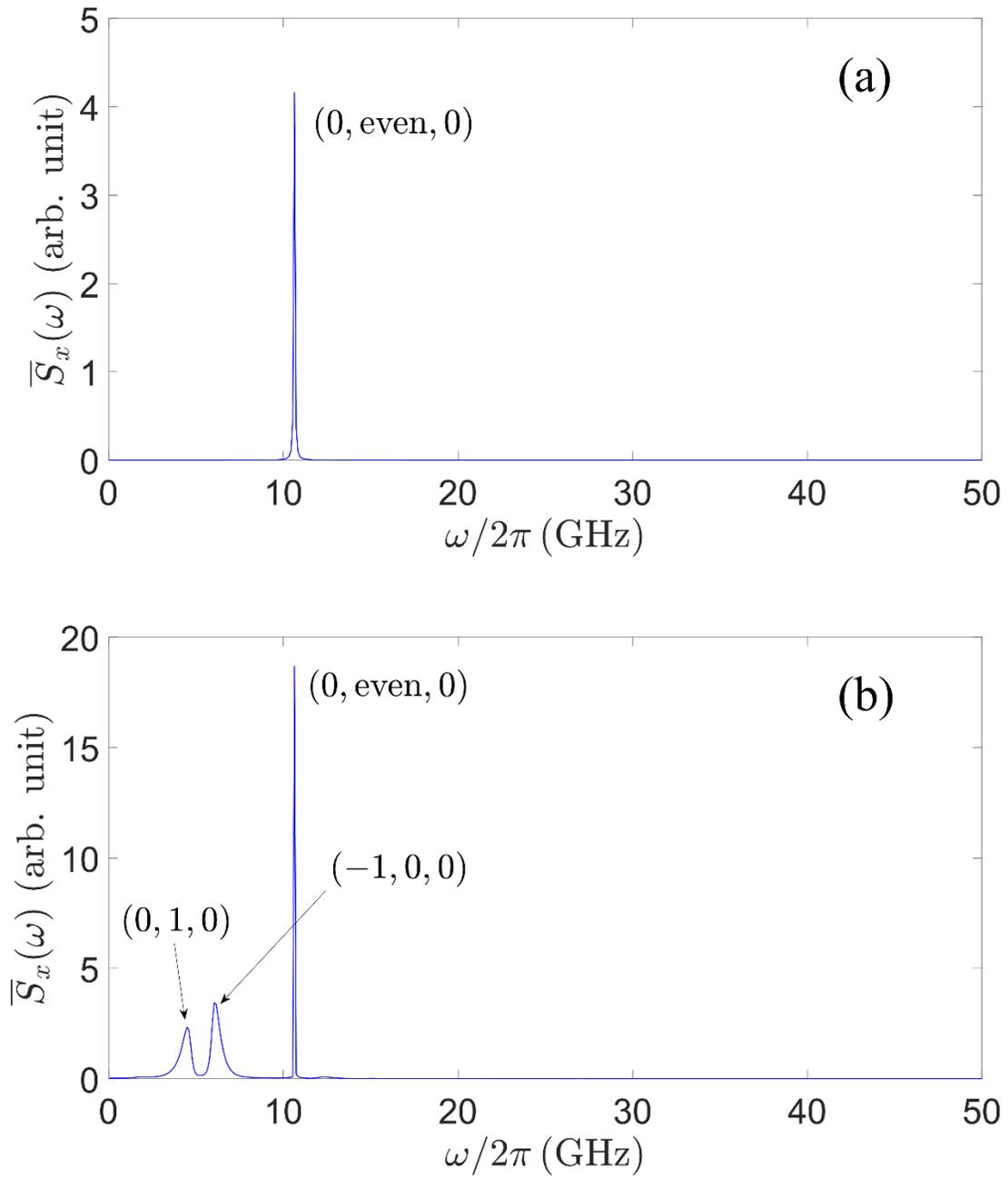

FIG. 2. Power spectral densities of the simulation in Fig. 1 extending over time windows (a) 0 to 10 ns (b) 30 to 40 ns.

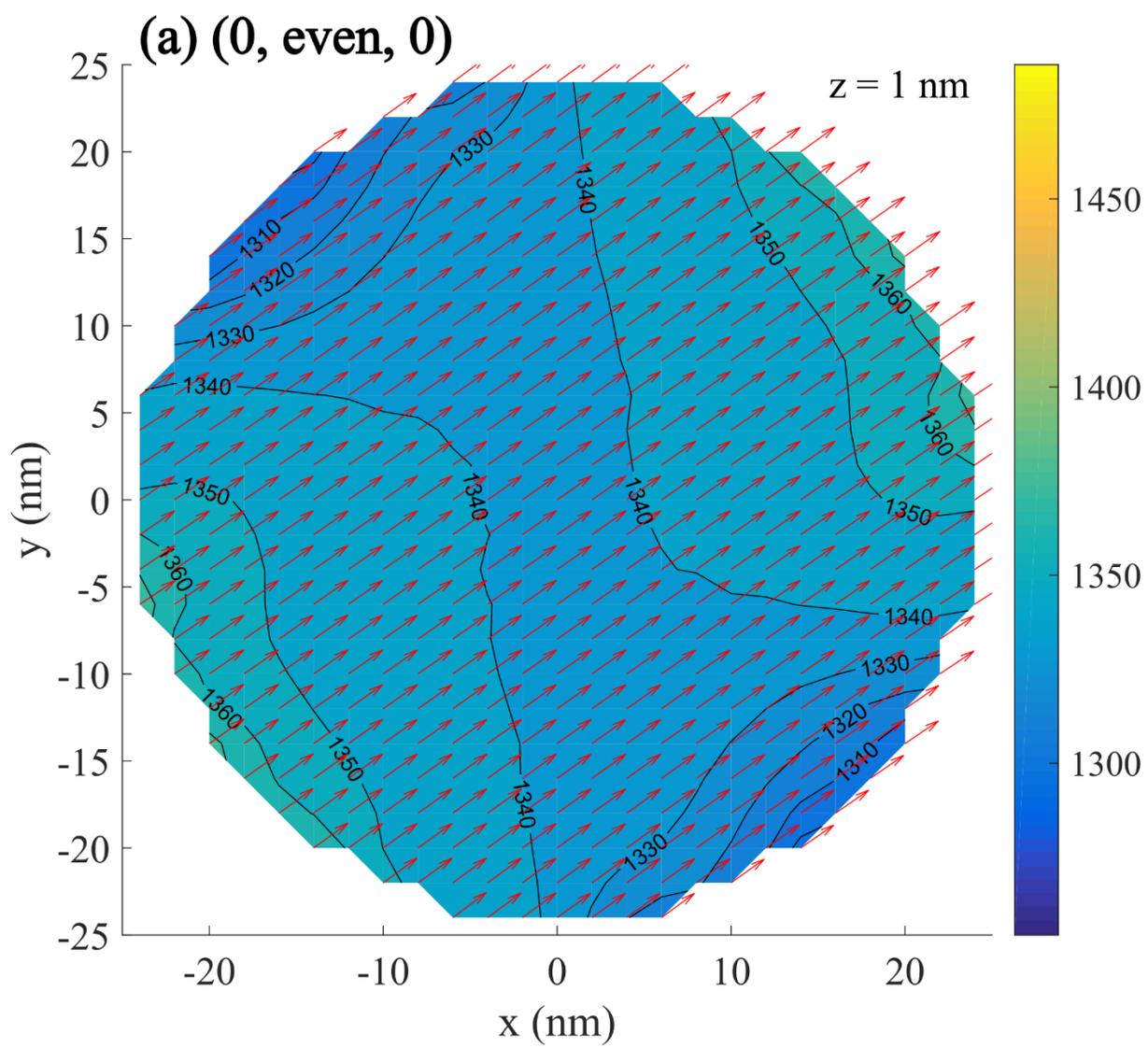

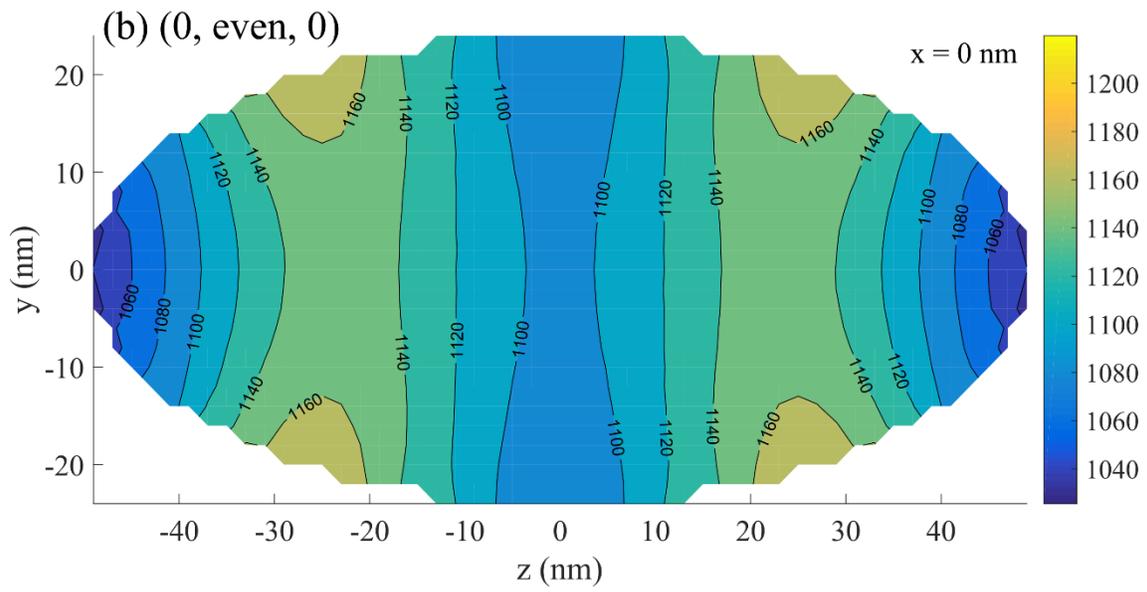

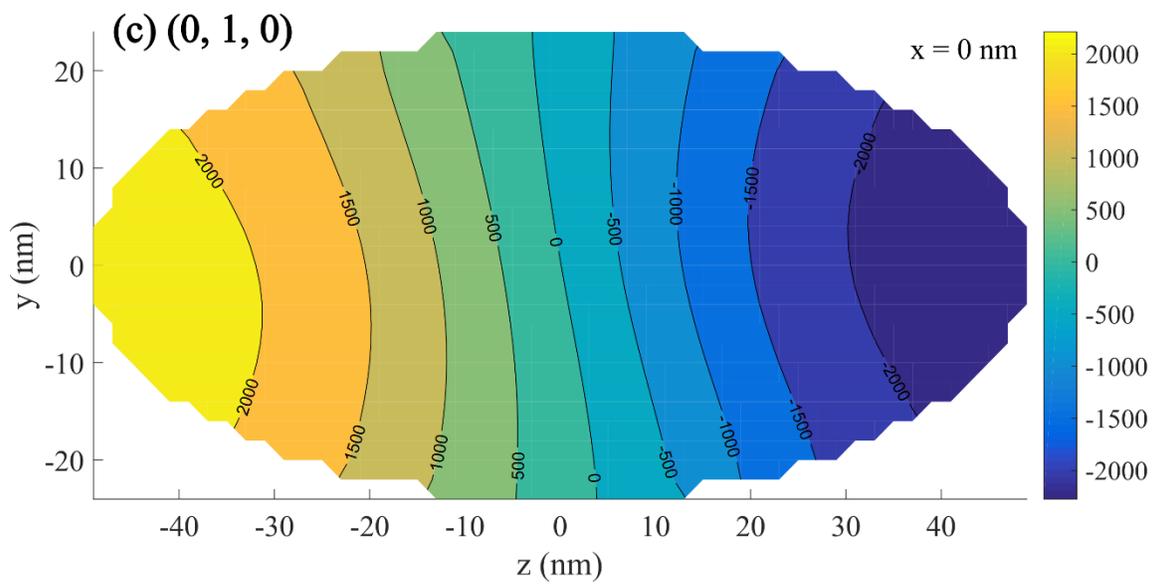

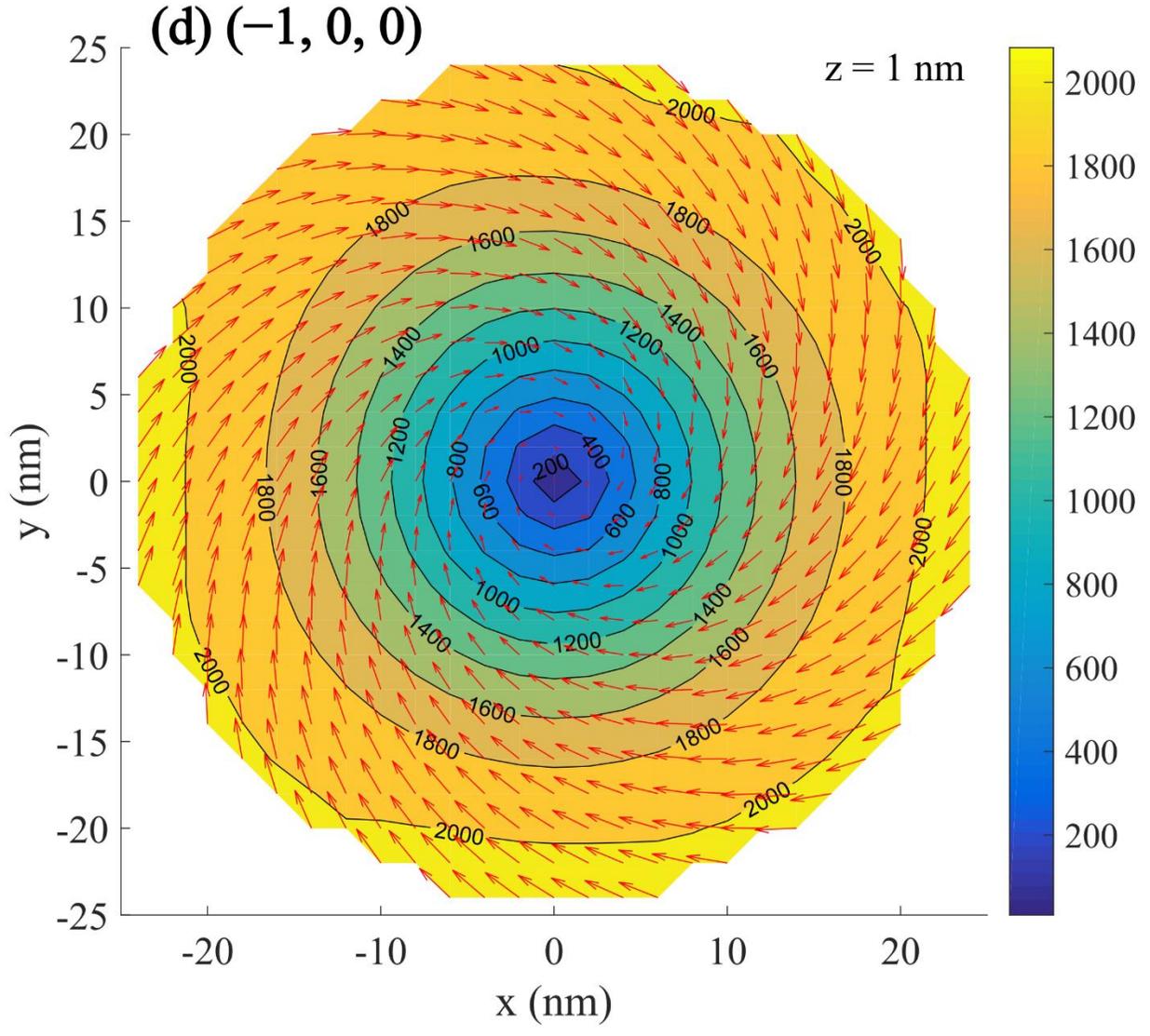

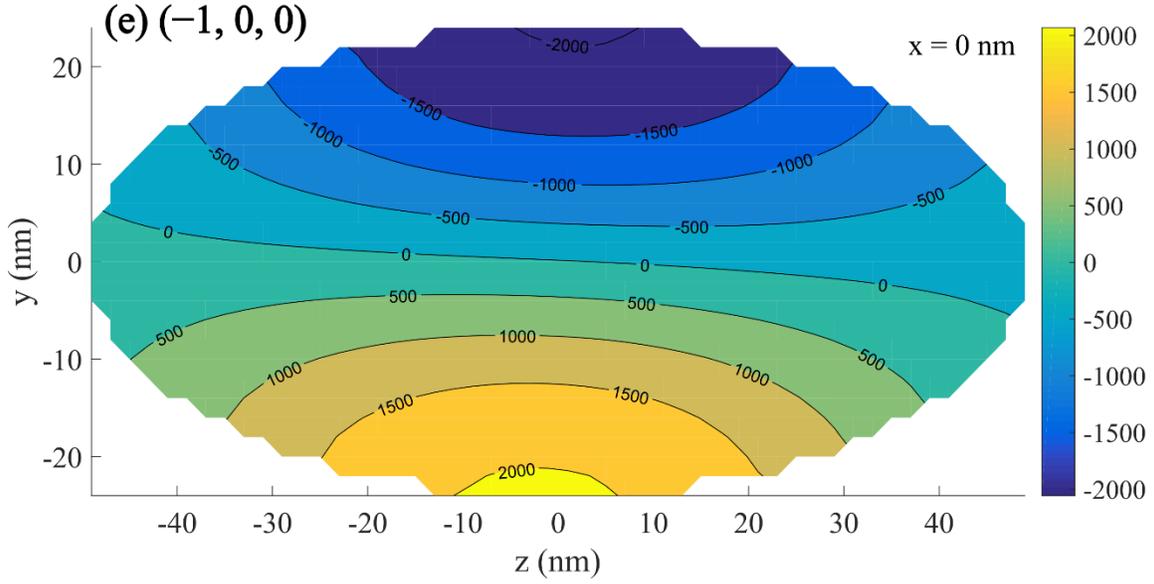

FIG. 3. Mode patterns associated with the simulation shown in Fig. 1 with $f_0 = 10.64$ GHz corresponding to Case 3 in Table II. (a) and (b): x-y and y-z cross sections of the real part of the mode amplitude of the (0, even, 0) mode during 0–10 ns. (c): similar y-z cross section for the (0, 1, 0) mode during 30–40 ns. (d) and (e): x-y cross section of the real part and y-z cross section of the imaginary part of the amplitude for the (−1, 0, 0) mode during 30–40 ns respectively. The x-y cross section of the (0, 1, 0) mode is omitted because it is very similar to part (a). The quantum numbers of the modes are immediately evident from these patterns.

To show that the microwave field (4.1) excites only modes within the group (0, even, 0), we conducted a simulation starting from the state at 10 ns of the simulation in Fig. 1, and turned off the microwave field and the damping. This allows the nonresonantly driven state to decay into normal modes. Fig. 4 shows a power spectral density and a y-z cross section of the (0, 0, 0) mode. Fig. 4(a) shows that the (0, 2, 0) peak on the shoulder of the main (0, 0, 0) peak is very small (less than 1% of the main peak) and not observable on this scale. By comparing Fig. 4(b) with Fig. 3(b), one can observe that the (0, even, 0) mode has a small contamination, even in z, corresponding to a small (0, 2, 0) mode amplitude. We have found that the relative strengths of the normal modes making up the weakly nonlinear (0, even, 0) excited mode depend on the excitation frequency, $f_0$.

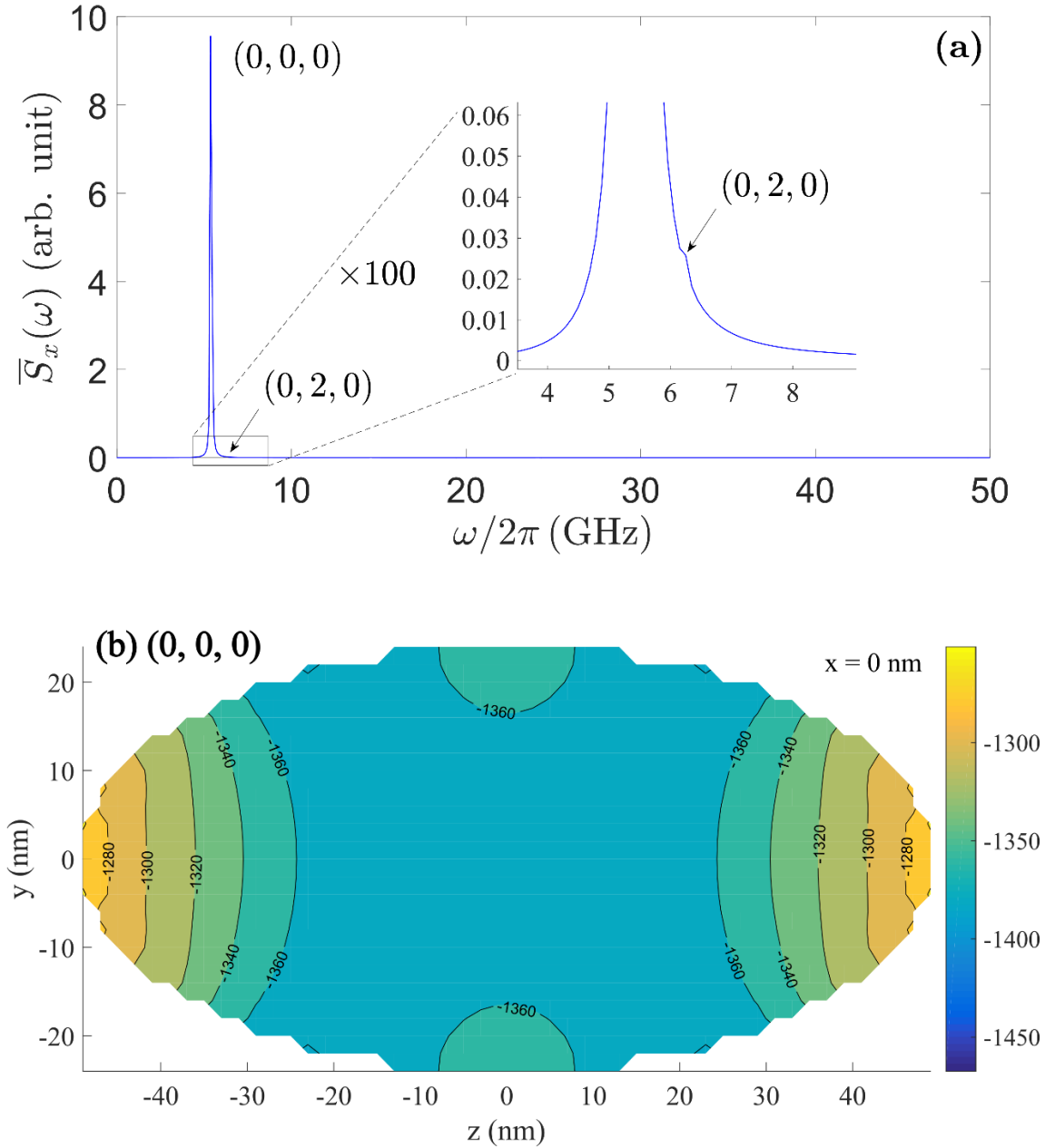

FIG. 4. Power spectrum (a) and y-z cross section of the (0, 0, 0) mode resulting from free (undriven, non-dissipative) evolution of the state at 10 ns in Fig. 1. The time window for the FFT is 10 ns.

The simulation just described shows a clear three-mode down conversion process. As already mentioned, it is listed as Case 3 in Table II. The table shows all the various three-mode down conversion processes we have examined. It lists $f_0 = \omega_0/2\pi$, followed by the mode numbers $(p_1, n_{z1})$ and $(p_1, n_{z2})$ of the final state modes; $n_{r1} = n_{r2} = 0$ for all cases. We then show the frequencies $f_1'$ and $f_2'$ of these modes along with their small amplitude (or linear regime) values $f_1$ and $f_2$ in parentheses. Not all the processes are observed. In many of these cases, we reran the simulation with a greater value of $\dot{H}_1$ or smaller value of the damping $\alpha$, or both.

We note here that an alternative to solving the Landau-Lifshitz partial integro-differential equation is to solve the ordinary differential equations for the mode amplitudes in a mode coupling formalism [79]. Such an approach is attractive and effective in many situations, but it is not without shortcomings. The number of modes that must be kept and the order to which they must be coupled is not known a priori. Further, we observe that the post-instability evolution seen for $t \gtrsim 40$ns in Fig. 1 is full of rich structure. We have studied this behavior in many ways using different protocols, and found many systematic effects, which we hope to report in a separate publication. Since $m_z$ can even become negative in this time development, it is highly dubious that is could be captured by a mode coupling treatment.

Table II. Observed and unobserved three-mode down conversion processes excited by an initial uniformly precessing state. The second column lists the quantum numbers $(p_1, n_{z1})$ and $(p_2, n_{z2})$ of the two final-state modes; $n_r = 0$ for all modes. All processes tabulated satisfy the selection rules (3.1)–(3.3).

| Case No. | Final state modes | | Frequencies (GHz): $f_{observed}$ ($f_{linear}$) | | | Observed? | Comments |
|---|---|---|---|---|---|---|---|
| | $(p_1, n_{z1})$ | $(p_2, n_{z2})$ | $f_0$ | $f_1$ | $f_2$ | | |
| 1 | (0, 0) | (−1, 1) | 15.72 | (5.66) | (11.23) | No | Not seen with $\alpha_{py}/80$ |
| 2 | (0, 0) | (−1, 3) | 27.00 | (5.66) | (21.88) | No | |
| 3 | (0, 1) | (−1, 0) | 10.06 | 4.40 (4.98) | 5.76 (6.45) | Yes | $\alpha = \alpha_{py}$; Tip ~5° |
| 4 | (0, 1) | (−1, 2) | 20.00 | (4.98) | (16.21) | No | Not seen with $\alpha_{py}/80$ |
| 5 | (0, 2) | (−1, 1) | 15.72 | 6.06 (6.64) | 9.67 (11.23) | Yes | $\alpha = \alpha_{py}$; Tip ~5° |
| 6 | (0, 2) | (−1, 3) | 27.00 | (6.64) | (21.88) | No | |
| 7 | (0, 3) | (−1, 0) | 15.72 | (10.06) | (6.445) | No | |
| 8 | (0, 3) | (−1, 2) | 20.50 | 7.62 (10.06) | 12.89 (16.21) | Yes | $\dot{H}_1 = 13.3$ Oe/ns, Tip ~19° |
| 9 | (1, 1) | (−2, 0) | 27.00 | 11.33 (14.36) | 15.63 (19.06) | Yes | Tip ~16° |
| 10 | (1, 2) | (−2, 1) | 38.50 | 16.41 (17.87) | 22.07 (23.93) | Yes | $\dot{H}_1 = 4$ Oe/ns (a default value for Table II) |
| 11 | (0, 0) | (−1, 5) | 41.00 | (5.66) | (35.94) | No | |

A. Discussion of individual cases

We now discuss the various cases listed in Table II one by one. Recall that our mode label is $(p, n_z, n_r)$, abbreviated to $(p, n_z)$ when $n_r = 0$. We find that the selection rules are always satisfied. The p selection rule requires that $p_{f1} + p_{f2} = -1$, which can be met by taking (a) $p_{f1} = 0$, $p_{f2} = -1$ (seen in cases 3, 5, and 8); or (b) $p_{f1} = 1$, $p_{f2} = -2$ (cases 9 and 10); or still higher values of $p_{f1} - p_{f2}$ (never seen). In all cases where an instability occurs, we find $n_{z1} - n_{z2} = 1$.

**Case 1.** We ran our basic simulation with $f_0 = 15.72$ GHz with the idea of seeing the process

$$(0, 0) \rightarrow (0, 0) + (-1, 1) \tag{4.3}$$

for which $f_1$ and $f_2$ are 5.66 and 11.23 GHz. These would appear to be the very lowest quantum number final states, but we failed to see an instability even with $\alpha = 10^{-4}$.

**Case 2.** In this case we chose $f_0 \simeq 27$ GHz, with the intent of looking for the final state $(0, 0) + (-1, 3)$, at frequencies 5.66 and 21.88 GHz. This process is technically allowed by the selection rules, but was not observed, even after the initial state was seeded with a small amount of the $(0, 0)$ and $(-1, 3)$ mode patterns amounting to about a 1° tipping. We were concerned that because these modes were being driven non-resonantly, they might damp out, and so we also tried to add the seed at an intermediate time, but this also did not drive the sought for process.

At very high amplitude ($m_\perp \simeq 0.28$, or a 16° tip), this drive showed an instability to $(1, 1) + (-2, 0)$ with pulled-down frequencies of 11.33 and 15.63 GHz. But, the mode patterns are not clean: the symmetry axes for both modes are not perfectly along $z$, and the $p = -2$ pattern looks a bit like two spatially separated $p = -1$ vortices. This is followed by chaotic behavior and finally undergoes magnetization reversal from $+\hat{\mathbf{z}}$ to $-\hat{\mathbf{z}}$. This is listed as Case 9 in Table II.

**Case 3.** The final state here is $(0, 1) + (-1, 0)$, at shifted frequencies 4.40 and 5.76 GHz, from linear regime values 4.98 and 6.45 GHz; $f_0$ varied from 10.06 to 10.64 GHz. The frequency shifts are associated with $m_\perp \simeq 0.1$.

**Case 4**. Here we sought (and failed) to see the final state (0, 1) + (−1, 2), for which

$$f_1 + f_2 = 4.98 + 16.21 = 21.19 \text{ GHz.} \tag{4.4}$$

We tried $f_0 = 20.5$ GHz, both with the default $\dot{H}_1 = 4$ Oe/ns, and then with 13.3 Oe/ns. In the latter case we saw an instability to (0, 3) + (−1, 2) with significantly pulled-down frequencies 7.62 and 12.89 GHz; the associated tip angle is ∼19° – this is listed as Case 8 in Table II.

We also tried $f_0 = 20.0$ GHz, and a much smaller damping, $\alpha = 10^{-4}$. Again, we only saw Case 8 with similar frequency pulling.

**Case 5**. Our initial intent was to look for the decay to (0, 2) + (0, 3), which would violate the *p* selection rule. Instead we found a final state (0, 2) + (−1, 1), with moderate frequency pulling (6.64 to 6.06 and 11.23 to 9.67 GHz). Note that the linear mode frequency sums for these two final states are quite close.

This same final state was observed with comparable frequency pulling in a simulation that was run in order to see the final state (0, 3) + (−1, 0) — listed as Case 7. Here, we chose $f_0 = 15.72$ GHz with the idea that since $f_{0,3} + f_{-1,0} = 16.51$ GHz is less than $f_{0,2} + f_{-1,1} = 17.87$ GHz, the (0, 3) + (−1, 0) instability should happen first and preempt the instability to (0, 2) + (−1, 3). This did not happen, so we tried adding seeds for the (0, 3) and (−1, 0) states, but this also produced the (0, 2) + (−1, 1) final state with similar frequency shifts (5.76 and 9.96 GHz).

Yet the same final state [(0, 2) + (−1, 1)] was observed (with frequencies 6.25 and 11.74 GHz) when we started with $f_0 = 17$ GHz with the goal of seeing the p-selection-rule-violating transition to (−1, 0) + (−1, 1).

**Case 6**. Final state (0, 2) + (−1, 3) [$f_1 + f_2 = 6.64 + 21.88 = 28.52$ GHz; $f_0 = 27.00$ GHz] — not observed. We comment on this case along with others in this family below.

**Case 7**. Final state (0, 3) + (−1, 0) — not observed; discussed in connection with Case 5.

**Case 8**. Here the final state is $(0, 3) + (−1, 2)$, which we have discussed in part in connection with Case 4. This state was also seen with much smaller frequency pulling in a run with $f_0 = 25.0$ GHz.

**Case 9**. Final state $(1, 1) + (−2, 0)$ — already discussed along with Case 2. Note that in this case, neither of the two final state modes has $p = 0$.

**Case 10**. Now the final state is $(1, 2) + (−2, 1)$. This is similar to Case 9 in that neither of the final state modes has $p = 1$. We chose $f_0 = 38.5$ GHz, and saw modest (~10%) frequency pulling ($f'_{1,2} = 16.4$ and $22.7$ GHz vs. $f_{1,2} = 17.9$ and $23.9$ GHz).

**Case 11**. Final state $(0, 0) + (−1, 5)$ — not observed; discussed in connection with Case 5.

## B. Discussion

Let us now offer some qualitative remarks about when a particular instability is seen or not seen. In general, for an instability to occur, the driving amplitude, $|H_1|$, must exceed a threshold value, $H_t$, (B.7). This threshold is large if the damping $\alpha$ is large, or if the coupling constant $g_{\alpha,\beta\gamma}$ [see Eq. (A.41)] is small. In addition, however, we must keep in mind that if $H_t$ is large, the modes $\beta$ and $\gamma$ into which the decay is being sought are perturbations on a background magnetization which is precessing at a large tipping angle from the z direction (see, e.g., Case 8), and their frequencies $f'_\beta$ and $f'_\gamma$ may be significantly detuned from the linear regime or normal mode values $f_\beta$ and $f_\gamma$. In the series expansion of the energy in terms of small tipped magnetization components as given at Eqs. (A.15–20), as the initial excitation amplitude grows one needs higher order terms describing three or more mode interactions. In such situations, the frequencies of other normal modes depend on the amplitude of the uniform mode. Further, the condition for decay is no longer Eq. (3.1) but

$$f_0 = f'_\beta + f'_\gamma. \tag{4.5}$$

Most of the time $f'_\beta < f_\beta$ and $f'_\gamma < f_\gamma$ as can be seen from Table II, and therefore, $f_0$ must generally be chosen to be smaller than the sum of the linear regime or small-tipping-angle frequencies.

Except for cases 9 and 10, the cases we have examined all have final states of the type

$$(0, n_{z1}) + (-1, n_{z2}). \tag{4.6}$$

As a rule, large values of either $n_{z1}$ or $n_{z2}$ are not seen even when the parity selection rule is obeyed. Presumably this is because the threshold $H_t$ is too high on account of a small coupling constant $g_{\alpha,\beta\gamma}$, and indeed, it is clear that the overlap integral (A.40) will be small if $F_\beta$ or $F_\gamma$ has many nodes in z. However, we only see cases where $n_{z1} = n_{z2} + 1$, and never see the case where $n_{z1} = n_{z2} - 1$. We do not have a proper explanation for this. That would require actual numerical evaluation of the relevant coupling constants. We caution that even such a calculation may not be adequate because of the frequency pulling effect mentioned in the previous paragraph and mode-dependent dampings discussed in Sec. III and Appendix B. Whether a perturbation with the appropriate angular and z symmetry can always be regarded as a continuous deformation of a small amplitude normal mode when the perturbation is on a large tipping angle background is unclear, and quantitative estimates of the detuning as a function of the tipping angle are likely to be difficult.

## V. Three-mode instabilities with $p_i = -1$, $n_{i,z}$ = even excitation

The initial study of the Suhl instability was motivated by experiments on large amplitude excitation of the uniform FMR mode, but simulations offer the opportunity to see how nonuniform FMR modes destabilize. In this section we consider the case $p_i = -1$.

We can generate fields that can excite modes with $p_i = -1$ using realistic coils as follows. Fig. 5(a) shows a pair of anti-Helmholtz coils the planes of which are perpendicular to the z-axis. The radii of both coils are $r_c$ and distance between them is $d_c$. We will assume that $r_c$ and $d_c$ are at least three to four times larger than the dimensions of our spheroidal sample. We further assume that equal, but oppositely directed, ac currents at microwave frequency flow in each coil. Then, provided $r_c$, $d_c \ll 2\pi c/\omega_0$ (c being the speed of light), the effects of retardation may be ignored, and in leading order the magnetic field near the center of the sample is given by the instantaneous magnetostatic form [80],

$$\mathbf{H}_{\text{mw}}(\mathbf{r},t) = h(x\hat{\mathbf{x}} + y\hat{\mathbf{y}} - 2z\hat{\mathbf{z}})\sin(\omega_0 t). \tag{5.1}$$

Such a field has the same azimuthal pattern as the magnetization of $p = -1$ modes (see e.g. Fig. 13a of [63]) so that one can expect it to excite some of these modes.

Fig. 5(b) describes a coil set up to excite $p = 1$ modes. We discuss this excitation in detail in Sec. VI.

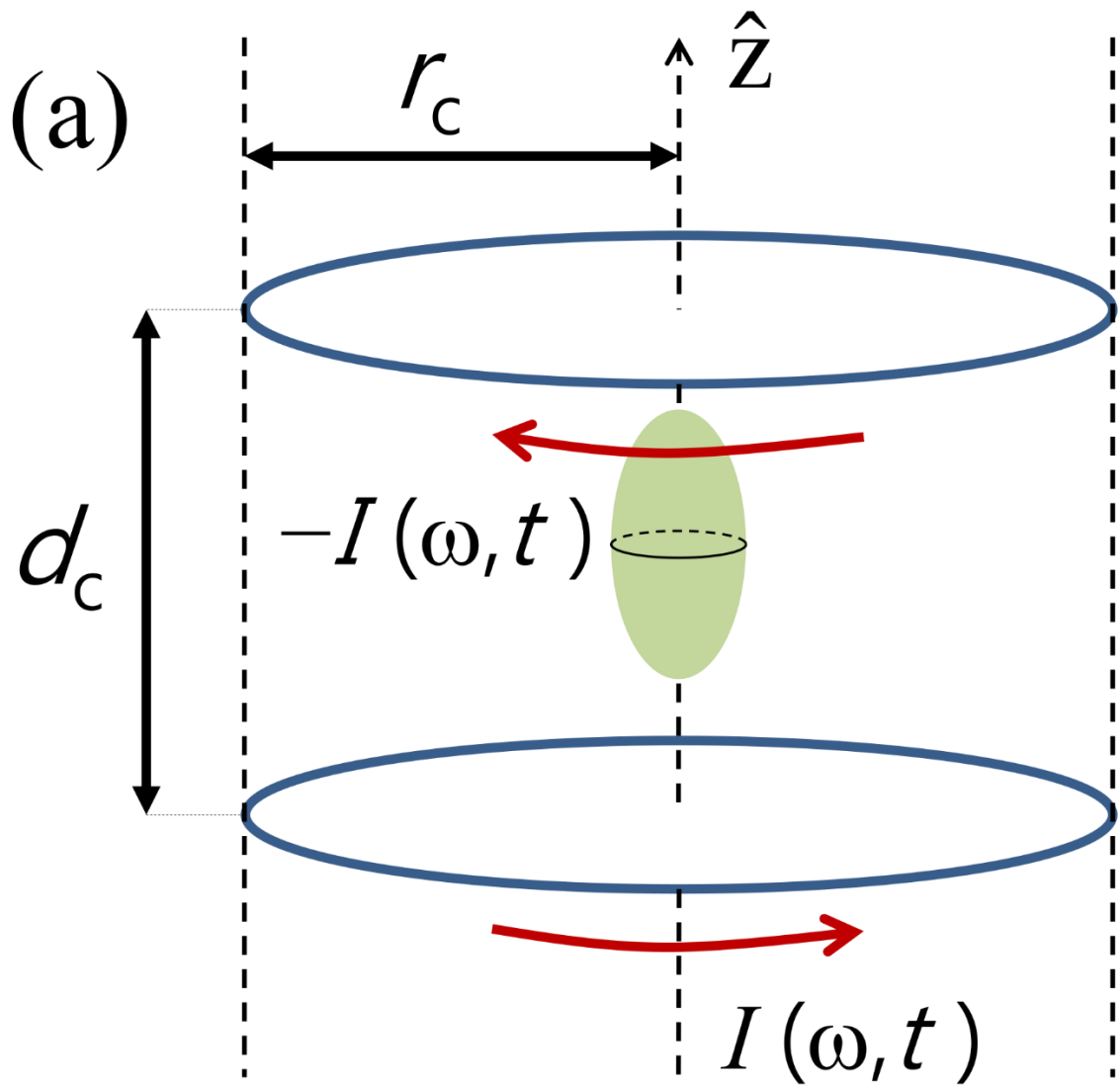

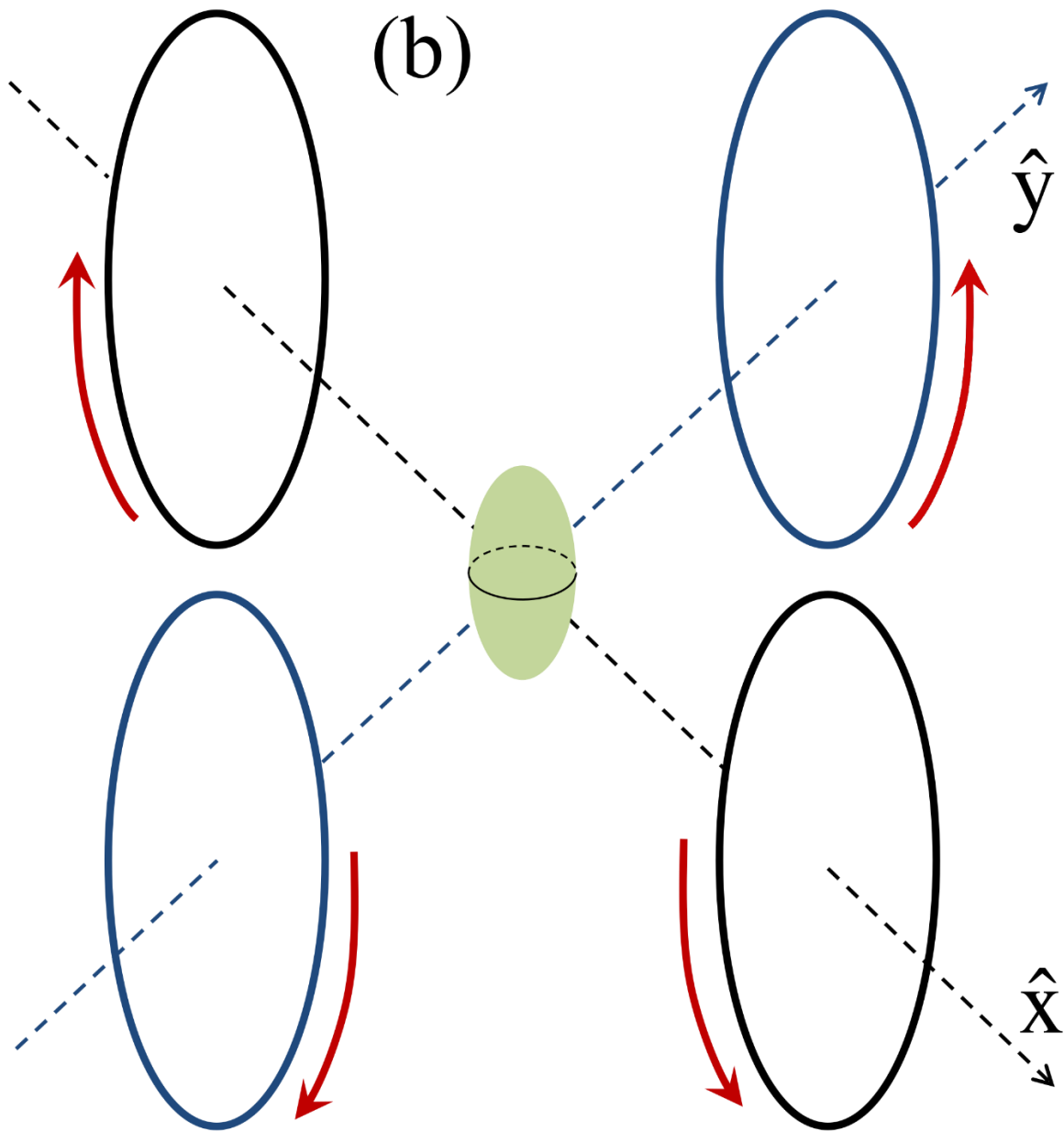

FIG. 5. Two different coil set ups; (a) to generate $\mathbf{H}_{\mathrm{mw}}(\mathbf{r},t)$ of Eq. (5.1); (b) to generate $\mathbf{H}_{\mathrm{mw}}(\mathbf{r},t)$ of Eq. (6.2).

In our simulations, we therefore used microwave fields of the form (5.1) with $h$ replaced by an amplitude $H_1(t)$ that is time dependent as in Sec. IV. Fig. 6(a) shows the PSD of an initial response before the onset of instabilities with $f_0 = 17$ GHz and $\dot{H}_1 = 8$ Oe/ns corresponding to Case 1 in Table III. This is a non-resonantly driven response and not a natural mode of the system. The FFT shows that this response consists of a single frequency that is the same as $\omega_0$. When one turns off the microwave field and the damping, this non-resonantly driven response decays into several natural modes of the system. This can be seen in Fig. 6(b) which shows the PSD for a 10 ns window after the drive field is turned off. One can see several peaks mainly consisting of $(p, n_z, n_r) = (-1, 0, 0), (-1, 2, 0)$, and $(-1, 4, 0)$. We note that the relative strengths of these modes depend on the excitation frequency. Fig. 7 shows the associated mode patterns which confirm the mode assignments. Fig. 7(a) shows that the initial driven mode definitely has an even profile in $z$. We also found that its xy-plane profile resembles $(p, n_r) = (-1, 0)$. The xy and yz cross sections of $(p, n_z, n_r) = (-1, 0, 0)$ are shown in Figs. 3(d) and 3(e). The yz cross sections of $(p, n_z, n_r) = (-1, 2, 0)$, and $(-1, 4, 0)$ are given at Figs. 7(b) and 7(c) respectively. We conclude that this microwave field excites p = −1 modes that are even in z. From Eq. (3.3), one can expect that the final mode pairs will have opposite parity in z in three-mode down conversion.

30In our simulations, we therefore used microwave fields of the form (5.1) with $h$ replaced by an amplitude $H_1(t)$ that is time dependent as in Sec. IV. Fig. 6(a) shows the PSD of an initial response before the onset of instabilities with $f_0 = 17$ GHz and $\dot{H}_1 = 8$ Oe/ns corresponding to Case 1 in Table III. This is a non-resonantly driven response and not a natural mode of the system. The FFT shows that this response consists of a single frequency that is the same as $\omega_0$. When one turns off the microwave field and the damping, this non-resonantly driven response decays into several natural modes of the system. This can be seen in Fig. 6(b) which shows the PSD for a 10 ns window after the drive field is turned off. One can see several peaks mainly consisting of $(p, n_z, n_r) = (-1, 0, 0), (-1, 2, 0)$, and $(-1, 4, 0)$. We note that the relative strengths of these modes depend on the excitation frequency. Fig. 7 shows the associated mode patterns which confirm the mode assignments. Fig. 7(a) shows that the initial driven mode definitely has an even profile in $z$. We also found that its xy-plane profile resembles $(p, n_r) = (-1, 0)$. The xy and yz cross sections of $(p, n_z, n_r) = (-1, 0, 0)$ are shown in Figs. 3(d) and 3(e). The yz cross sections of $(p, n_z, n_r) = (-1, 2, 0)$, and $(-1, 4, 0)$ are given at Figs. 7(b) and 7(c) respectively. We conclude that this microwave field excites p = −1 modes that are even in z. From Eq. (3.3), one can expect that the final mode pairs will have opposite parity in z in three-mode down conversion.

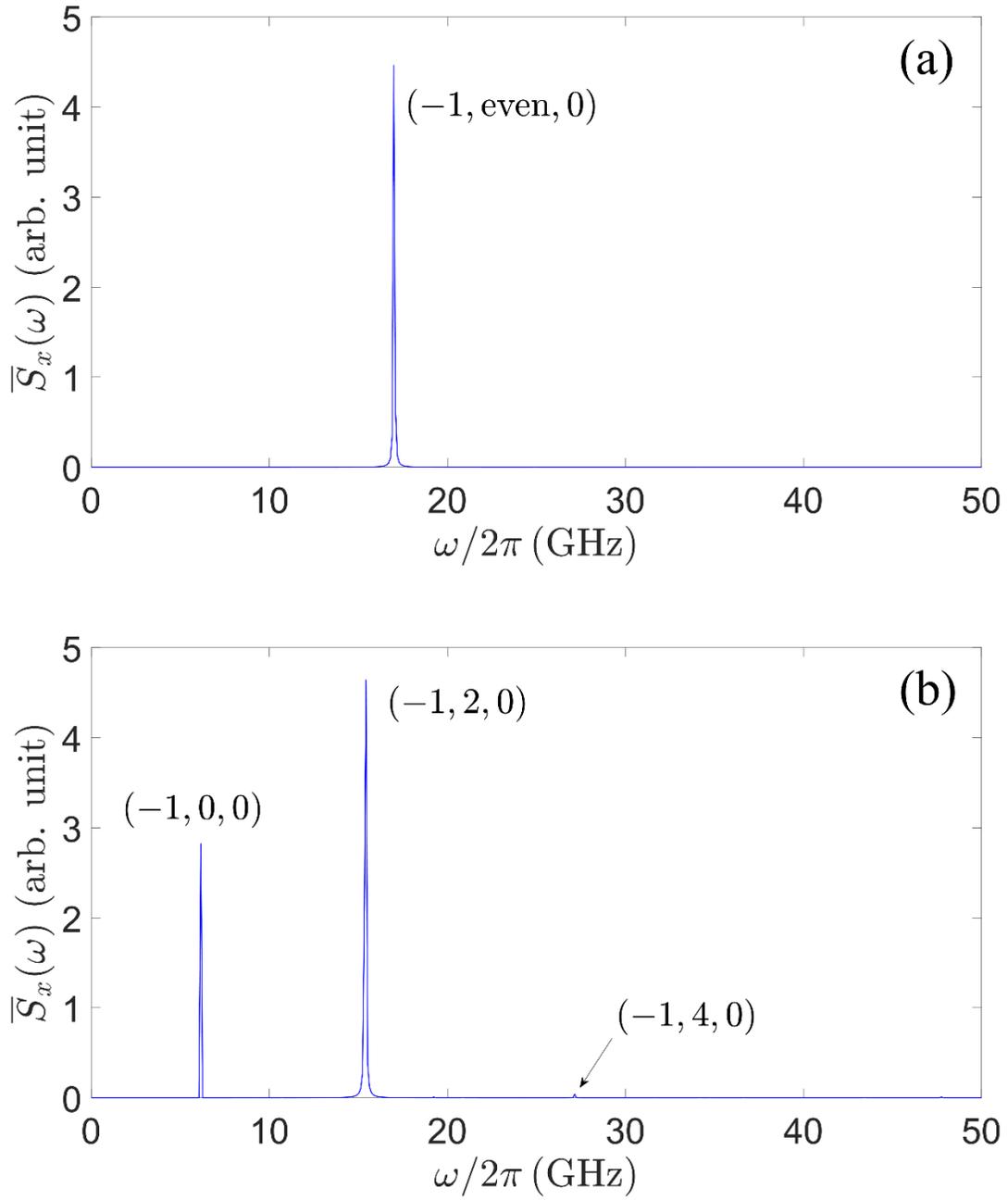

FIG. 6. Power spectral densities of (a) initial nonresonantly driven excitation of a simulation with a microwave field with $p = -1$ symmetry at $f_0 = 17$ GHz; (b) the state that develops from that generated in (a) if the driving field and damping are turned off.

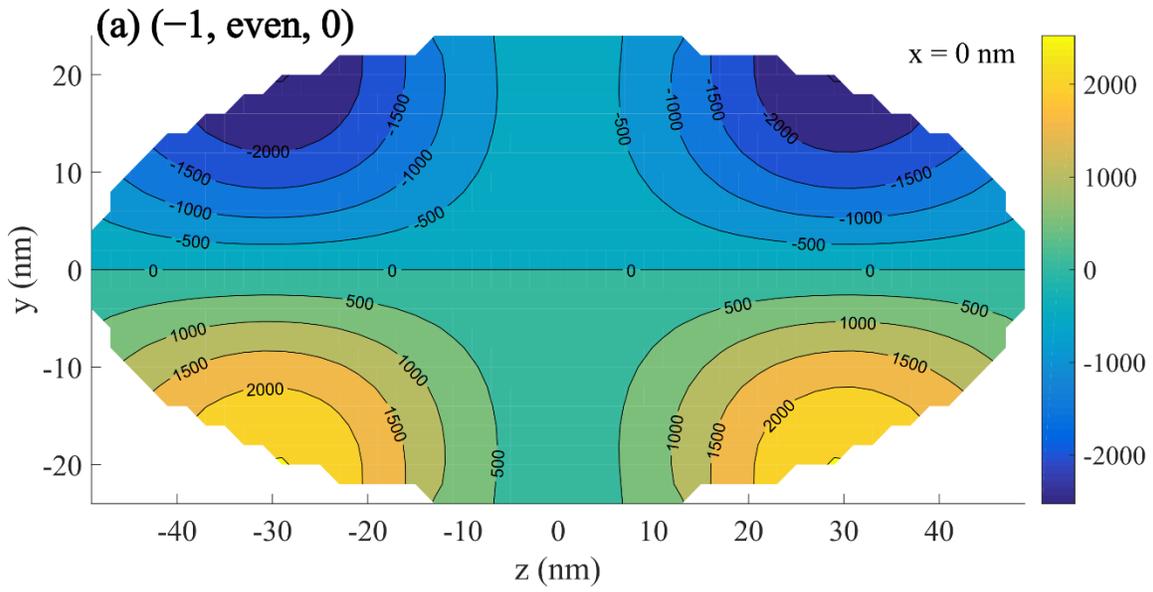

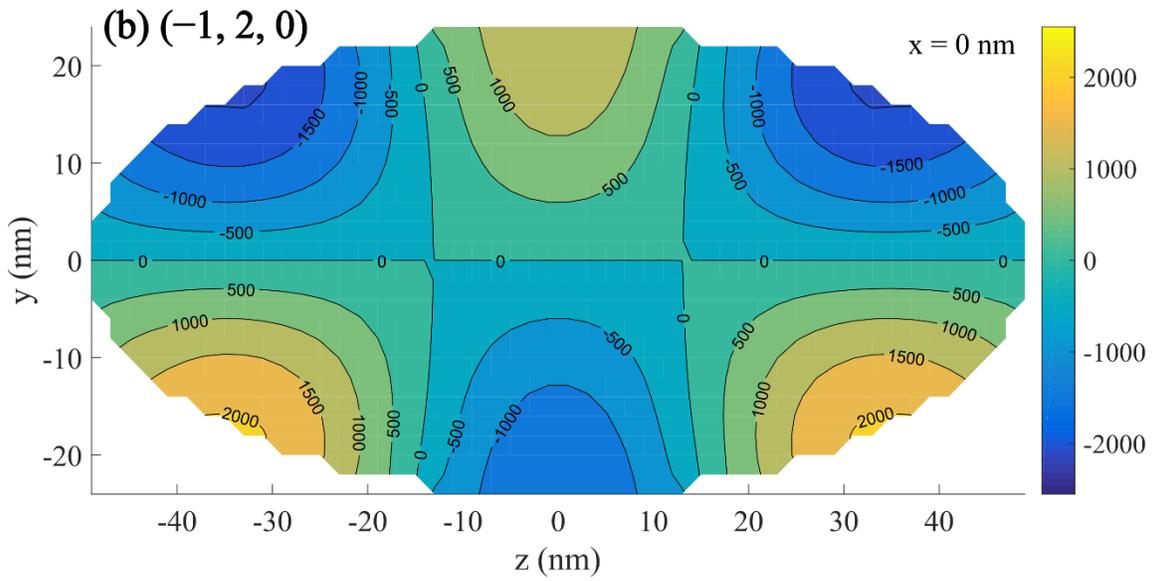

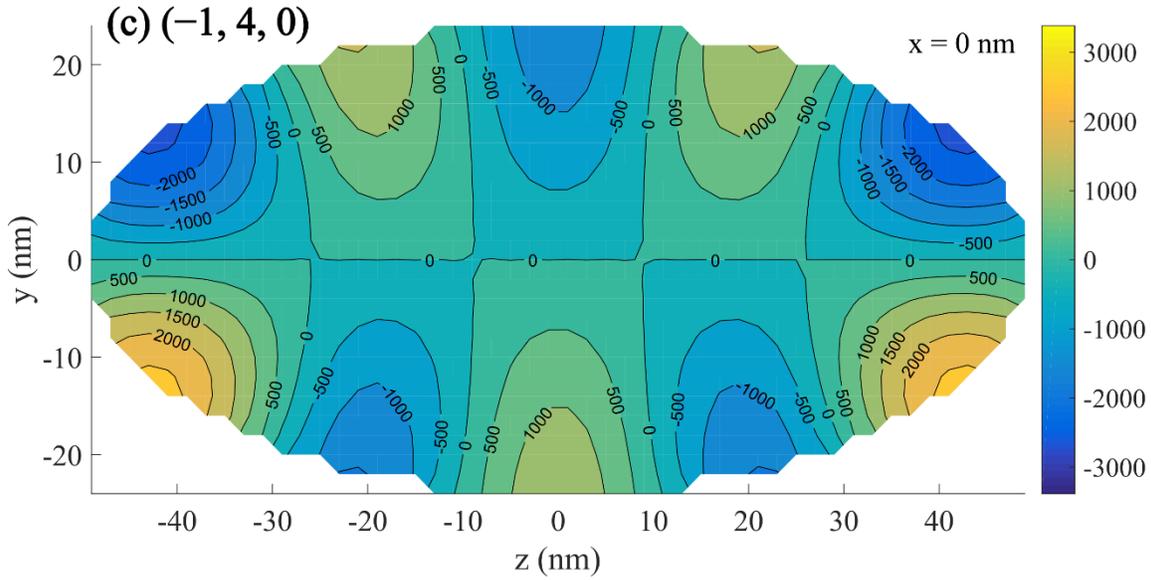

FIG. 7. Mode patterns, in yz-plane, associated to PSDs in Fig. 6: (a) imaginary part of the initial excitation, (b) and (c) real parts of the (−1, 4, 0) and (−1, 4, 0) modes in the subsequent free evolution.

The excitation of the sample and the analysis of the response follows the same path as in Sec. IV. As an example, Fig. 8 shows various plots for an instability corresponding to Case 1 in Table III with $f_0 = 17$ GHz and $\dot{H}_1 = 8$ Oe/ns. In Fig. 8(a) $M_x$ doesn't have noticeable changes because all initial and final modes have p = −1 in which the x and y components add to zero when we sum over the entire volume. By investigating the PSD and associated mode patterns, which are partially shown at Figs. 8(b) and (c), we identified which modes are excited at the instability. The patterns of the (−1, 0, 0) mode can be seen in Figs. 3(d) and 3(e).

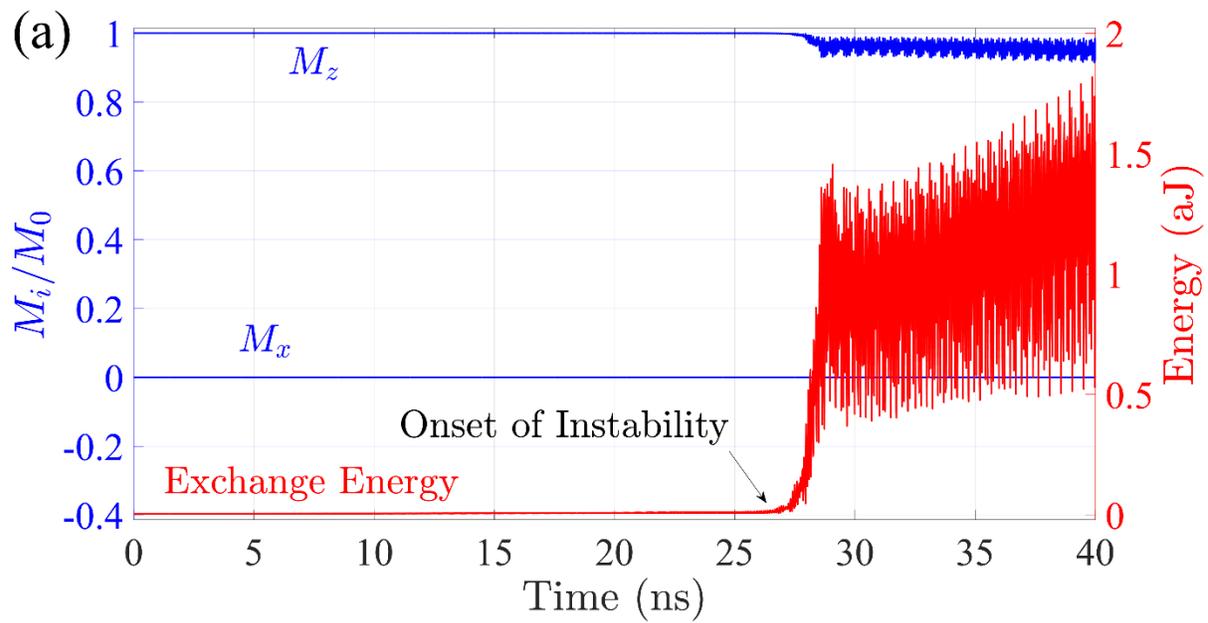

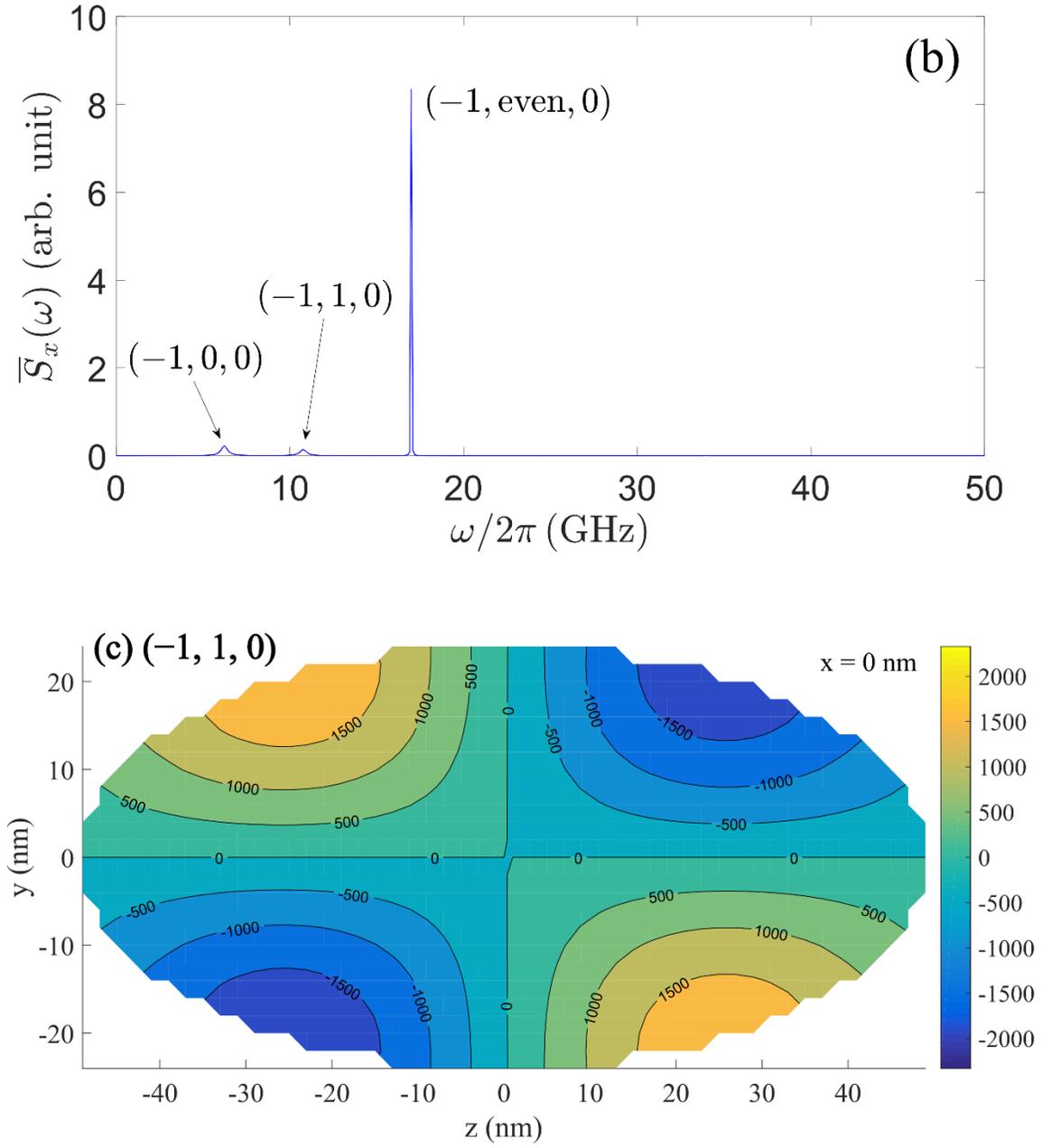

FIG. 8. A number of plots associated with the instability of Case 1 in Table III at $f_0 = 17$ GHz: (a) raw data which shows the onset of instability around 26 ns, (b) PSD for the 17–27 ns window, and (c) y-z cross section of the real part of the $(-1, 1, 0)$ mode amplitude.

Table III shows the various three-mode processes we have examined. The format is the same as in Table II. The selection rules (3.1)–(3.3) are satisfied for all three-mode processes we have examined as in the case of Table II. With ordinary permalloy damping, no instabilities associated with $p_1 = 0$ and $p_2 = -2$ (cases 5–12) were observed even if such combinations were allowed by the selection rules. We therefore conducted various lower damping simulations and observed an instability associated with ($p_1 = 0$, $n_{z1} = 1$) and ($p_2 = -2$, $n_{z1} = 0$) (Case 7) at two orders of magnitude lower damping ($\alpha = 1 \times 10^{-4}$). Since thresholds of Suhl instabilities depend on damping, this observation indicates that this instability has a much higher threshold than the others in Table III. This suggests that otherwise unobserved instability combinations were not due to a hidden selection rule, but due to a substantial difference in thresholds between allowed combinations. There are also some trends that one can find from Table II and III. Those are, first, that smaller $|n_{z1} - n_{z2}|$ is preferred, and second, having one or both of $p_1$ and $p_2$ be the same as the initial $p$ value, $p_i$, is preferred.

Table III. Observed and unobserved three-mode down conversion processes excited by an initial nonuniformly precessing state which can excite $p = -1$ modes. The first column lists the quantum numbers $(p_1, n_{z1})$ and $(p_2, n_{z2})$ of the two final-state modes; $n_r = 0$ for all modes. All processes tabulated satisfy the selection rules (3.1)–(3.3).

| Case No. | Final state modes | | Frequencies (GHz): $f_{observed}$ ($f_{linear}$) | | | Observed? | Comments |
|---|---|---|---|---|---|---|---|
| | $(p_1, n_{z1})$ | $(p_2, n_{z2})$ | $f_0$ | $f_1$ | $f_2$ | | |
| 1 | (−1, 0) | (−1, 1) | 17.00 | 6.25 (6.45) | 10.74 (11.23) | Yes | Complex process with $f_0 = 22$ GHz |
| 2 | (−1, 0) | (−1, 3) | 28.00 | (6.45) | (21.88) | No | |
| 3 | (−1, 1) | (−1, 2) | 28.00 | 11.72 (11.23) | 16.31 (16.21) | Yes | Complex process at 27 GHz and 6.67 Oe/ns |
| 4 | (−1, 2) | (−1, 3) | 39.50 | 17.19 (16.21) | 22.27 (21.88) | Yes | |
| 5 | (0, 0) | (−2, 1) | 29.50 | (5.66) | (23.93) | No | |
| 6 | (0, 0) | (−2, 3) | 42.00 | (5.66) | (36.43) | No | |
| 7 | (0, 1) | (−2, 0) | 23.00 | 4.88 (4.98) | 18.07 (19.04) | Yes | $\alpha = 1\times 10^{-4}$ |
| 8 | (0, 1) | (−2, 2) | 34.50 | (4.98) | (29.59) | No | |
| 9 | (0, 2) | (−2, 1) | 29.50 | (6.64) | (23.93) | No | |
| 10 | (0, 2) | (−2, 3) | 42.00 | (6.64) | (36.43) | No | |
| 11 | (0, 3) | (−2, 0) | 29.00 | (10.06) | (19.04) | No | |
| 12 | (0, 3) | (−2, 2) | 39.00 | (10.06) | (29.59) | No | |
| 13 | (−1, 0) | (−1, 5) | 42.00 | (6.45) | (35.94) | No | |
| 14 | (−1, 1) | (−1, 4) | 39.00 | (11.23) | (28.52) | No | |

A. Discussion of individual cases

We now discuss some of the cases listed in Table III individually, focusing mainly on those where an instability was seen. Note that now we must have $p_{f1} + p_{f2} = -2$. This can be satisfied via (a) $p_{f1} = p_{f2} = -1$ or (b) $p_{f1} = 0$, $p_{f2} = -2$ (or other combinations with still greater difference $p_{f1} - p_{f2}$). We find that the combination (b) is rare and seen only in one case (Case 7) below.

**Case 1**. The driving frequency $f_0 = 17.0$ GHz, and final state is $(-1, 0) + (-1, 1)$, which we see at $f'_{1,2} = 6.25$ and 10.74 GHz. By comparison, the linear regime frequencies are 6.45 and 11.23 GHz. Thus, both $f'_{1,2}$ are moderately downshifted (3-4%).

We also saw this case in a run with $f_0 = 22$ GHz and significantly upshifted frequencies $f'_{1,2} = 9.08$ and 12.89 GHz. Since the initial state is not a normal mode, we surmise that the final state here is a non-resonantly driven steady state with an observed average tipping angle of 4–5°. It seems that when the system doesn't have natural mode frequencies that obey the selection rules, sometimes the initial mode decays to final modes for which the unshifted frequency sum $f_1 + f_2$ is rather far from $f_0$ (of course, this case would have a higher threshold [4]), but even in that case the generalized frequency selection rule, (4.5), is well obeyed (for a big downshift case, see an analysis of Case 3 in Sec. V A). Very similar frequencies are seen in a run at 23 GHz, but the associated spin maps are not purely even or odd in z. It should be noted that the frequency here nearly matches the sum for the final state $(-1, 0) + (-1, 2)$. This state violates the parity rule and is not seen.

Staying with this case, we redid the run with $f_0 = 22$ GHz but a much lower damping, $\alpha = 10^{-4}$. We saw a final state with six modes: the initial driven mode at 22 GHz; the pair $(-1, 0) + (-1, 1)$ at the same frequencies as in the higher $\alpha$ run; an *upconversion* to a 44 GHz mode with $p$

= −1 and $n_z$ odd but mixed; a 31.05 GHz mode with $p = -1$ and $n_z = 3$; and a 34.86 GHz mode with $p = -1$ and $n_z$ even but mixed. This suggests that the upconversion process is

$$2 \times (-1, \text{even}) \rightarrow (-1, \text{odd}). \tag{5.2}$$

This is in accord with the selection rules. For comparison, we note that $f_{(-1,5)}$ and $f_{(-1,6)}$ are 35.94 and 45.61 GHz, so the upconverted state is some nonresonantly driven (−1, odd) mode. The peaks at 31.05 and 34.86 GHz suggest that we are seeing two subsequent down conversion processes

$$(-1, \text{odd}) \rightarrow (-1, 0) + (-1, \text{even}), \quad \text{and} \quad (-1, \text{odd}) \rightarrow (-1, 1) + (-1, 3). \tag{5.3}$$

These processes are (odd) → (even) + (even) and (odd) → (odd) + (odd) which are both allowed by (3.3). They cannot be generated directly via a three-mode instability from our initial state because that state has an even symmetry. This final pattern goes unstable very quickly to a further chaotic state.

**Case 2**. In a simulation that was run at $f_0 = 28$ GHz, we sought and failed to see the final state (−1, 0) + (−1, 3), for which $f_1 + f_2 = 28.33$ GHz. Instead, we saw Case 3. However, we saw the Case 2 final state in what we think is a four-magnon process, as we discuss in connection with Case 3.

**Case 3**. At $f_0 = 28.00$ GHz, we saw the final state (−1, 1) + (−1, 2), for which $f_1 + f_2 = 11.23 + 16.21 = 27.44$ GHz. We see $f'_{1,2} = 11.72$ and 16.31 GHz both of which are slightly upshifted. This case was also seen in a run at 27.50 GHz.

To examine the upshift phenomenon, we ran a simulation with $f_0 = 27$ GHz and $\dot{H}_1 = 6.67$ Oe/ns. We did see an instability to (−1, 1) + (−1, 2) with slight downshifts at about 100 ns, but this was followed some time (~2 ns) later by an upconverted mode at 54 GHz (aliased to 46 GHz since our Nyquist frequency is 50 GHz) with $p = -1$, $n_z$ odd but mixed, plus the following modes at the listed frequencies (all in GHz):

$$
\begin{aligned}
&\text{A: } (-1, 0) \text{ at } 6.54 &&\text{B: } (-1, \text{even}, n_r = 0 \text{ \& } 1) \text{ at } 47.46 \\
&\text{C: } (-1, 1) \text{ at } 11.23 &&\text{D: } (-1, \text{odd}) \text{ at } 42.77 \\
&\text{E: } (-1, 2) \text{ at } 15.82 &&\text{F: } (-1, \text{even}) \text{ at } 38.18 &&(5.4) \\
&\text{G: } (-1, 3) \text{ at } 20.41 &&\text{H: } (-1, 5) \text{ at } 33.69
\end{aligned}
$$

The maximum tipping angle of this multipeak magnetization pattern is about 12°. Our interpretation is that the upconverted mode at 54 GHz is subsequently undergoing two-mode decay to the pairs A + B, C + D, E + F, and G + H. Note that the selection rules are obeyed in each case. We speculate A + G (Case 2) is connected to the initial response at 27 GHz through a three-mode instability because this case also obeys the selection rules. Some mode patterns don't have a definite quantum number; D and F don't have definite $n_z$, and B doesn't have definite $n_r$. The pattern is followed by chaotic behavior at about 120 ns.

To study this case further, we then did a run at 28GHz with $\alpha = 10^{-4}$. This time the FFT shows four peaks just as in the low damping run for Case 1: a peak at 28 GHz, the expected $(-1, 1)$ and $(-1, 2)$ peaks, and an upconverted peak at 56 GHz (aliased to 44 GHz) whose mode pattern was $p = -1$, $n_z = 1$, $n_r = 1$. This is a rare case where a mode with $n_r \neq 0$ was excited.

The same case was seen in a run at 29 GHz where the goal was to see Case 11, and in a run at 24.50 GHz. In the last case, the frequencies are downshifted by about 2 GHz, suggesting that it is possible to drive the modes nonresonantly at a range of frequencies and tipping angles.

**Cases 4 and 14**. The final state in Case 4, which is observed, is $(-1, 2) + (-1, 3)$. As Table III shows, the frequencies are again upshifted. Both $(-1, 2)$ and $(-1, 3)$ have clear patterns at $f_0 = 39.50$ GHz, but $(-1, \text{odd}, n_r = \text{mixed})$ is observed instead of $(-1, 3)$ at $f_0 = 39.00$ GHz. Note that the linear frequency sum here is ~ 38.09GHz, which is close to Case 14, $(-1, 1) + (-1, 4)$ for which $f_1 + f_2 = 39.75$ GHz. Case 14 is not seen.

We also saw Case 4 with a driving frequency of $f_0 = 38$ GHz with frequencies very close to their linear regime values. This is a very clean case, perhaps the cleanest. We saw this case yet again in a run intended to see Case 8, with $f_0 = 34.50$ GHz. The frequencies $f'_{1,2}$ were downshifted from $f_{1,2}$ by 1–2 GHz.

**Case 7**. We first did a simulation at $f_0 = 23.50$ GHz with the goal of seeing the pair $(0, 1) + (-2, 0)$ with $f_{1,2} = 4.98$ and $19.04$ GHz. Instead we saw the Case 3 pair, $(-1, 1) + (-1, 2)$, at greatly downshifted frequencies $9.96$ and $13.96$ GHz.

We failed to see the Case 7 pair even after seeding the initial state with a small amount of the $(-2, 0)$ mode.

We then tried a run at 23 GHz with $\alpha = 10^{-4}$ and $\dot{H}_1 = 26.7$ Oe/ns, and succeeded in seeing the Case 7 pair at $f'_{1,2} = 4.88$ and $18.07$ GHz. However, at about 70 ns the system develops a very rich pattern which lasts about 10 ns, and the FFT also reveals an upconverted $(-1, \sim 5)$ mode at 46 GHz, plus the following modes at the listed frequencies (all in GHz):

A: $(0, 1)$ at $4.88$    B: $(-2, 1)$ at $41.11$    C: $(0, 0, n_r = 1)$ at $27.83$

D: $(-2, 0)$ at $18.07$    E: $(1, 1)$ at $9.77$    F: $(-3, 1)$ at $36.04$ (5.5)

The typical tipping angle in this multipeak pattern is about $2.5°$. Our interpretation (similar to Case 3) is that the upconverted mode at 46 GHz is undergoing two-mode decay to the pairs A + B, C + D, and E + F. The selection rules are obeyed in each case. The mode patterns are not all clean; the symmetry axis for mode B is not perfectly along z, and the $p = -2$ behavior looks a bit like two spatially separated $p = -1$ vortices. This pattern is followed by chaotic behavior at about 90 ns.

To summarize, when we start with a $p_i = -1$ excitation, we see a direct three-mode instability in only four cases (cases 1, 3, 4, and 7). Of these four, three are in the family

$$(-1, n_{z1}) + (-1, n_{z1} + 1). \tag{5.6}$$

We tried to look for many cases in which the final states had $p_{f1} = 0$ and $p_{f2} = 2$, and succeeded only once (Case 7).

## VI. Three-mode instabilities with $p_i = 1$, $n_{i,z}$ = even excitation

In this section we study three-mode processes when the initial excitation has $p = 1$. For this, we must first discuss how one can realistically make a time-dependent magnetic field which can couple to $p = 1$ modes. In contrast to the one pair of coils used to excite $p_i = -1$ modes, now we will use two pairs of anti-Helmholtz coils. The plane of the first pair will be perpendicular to the x-axis and that of the second pair will be perpendicular to the y-axis (see Fig. 5(b)). If we use the same approximations which we used for $p_i = -1$ and choose a proper relative phase difference between these pairs, the generated magnetic field has the spatial form

$$(x\hat{\mathbf{x}} + z\hat{\mathbf{z}} - 2y\hat{\mathbf{y}}) - (y\hat{\mathbf{y}} + z\hat{\mathbf{z}} - 2x\hat{\mathbf{x}}) = 3(x\hat{\mathbf{x}} - y\hat{\mathbf{y}}), \tag{6.1}$$

which has $p = 1$ symmetry. We therefore apply in our simulations a magnetic field

$$\mathbf{H}_{\mathrm{mw}}(\mathbf{r},t) = H_1(t)(x\hat{\mathbf{x}} - y\hat{\mathbf{y}})\sin(\omega_0 t), \tag{6.2}$$

with $H_1(t) = \dot{H}_1 t$ as before.

Fig. 9(a) shows a PSD of an initial response before the onset of instabilities with a microwave field of (6.2), $f_0 = 20$ GHz and $\dot{H}_1(t) = 13.3$ Oe/ns corresponding to Case 3 in Table IV. This is, as in Secs. IV and V, a non-resonantly driven response and not a natural mode of the system. A Fourier analysis shows that this response consists of a single frequency that is the same as $\omega_0$. When one turns off the microwave field, this non-resonantly driven response decays into several natural modes of the system. Fig. 9(b) shows a PSD over a 10 ns period collected after turning off the driving field. It consists mainly of $(p, n_z) = (1, 0)$, $(1, 2)$, and $(1, 4)$ modes. The fact that these resolved $p = 1$ modes are all even in z corresponds to what one can see in the actual magnetization initial response. Fig. 10 shows the associated mode patterns of the simulation used for Fig. 9. Figs. 10(a) and (b) shows x-y and y-z cross section of real component of (1, even, 0) mode function respectively; one can see that its xy-plane profile resembles a $p = 1$, $n_r = 0$ mode and its pattern along z-axis is even. Figs. 10(c), (d), and (e) shows y-z cross section of (1, 0, 0), (1, 2, 0), and (1, 4, 0) mode functions respectively; by comparing these with y-z cross section of $p = -1$ modes with the same $n_z$, one can notice that y-z cross sections of $p = \pm 1$ modes are actually the same. A detailed description of mode patterns in cylindrical symmetric samples can be found in Lim et al. [63]. Again, one can expect that two final states will be an even and odd pair in a three-mode down conversion case.

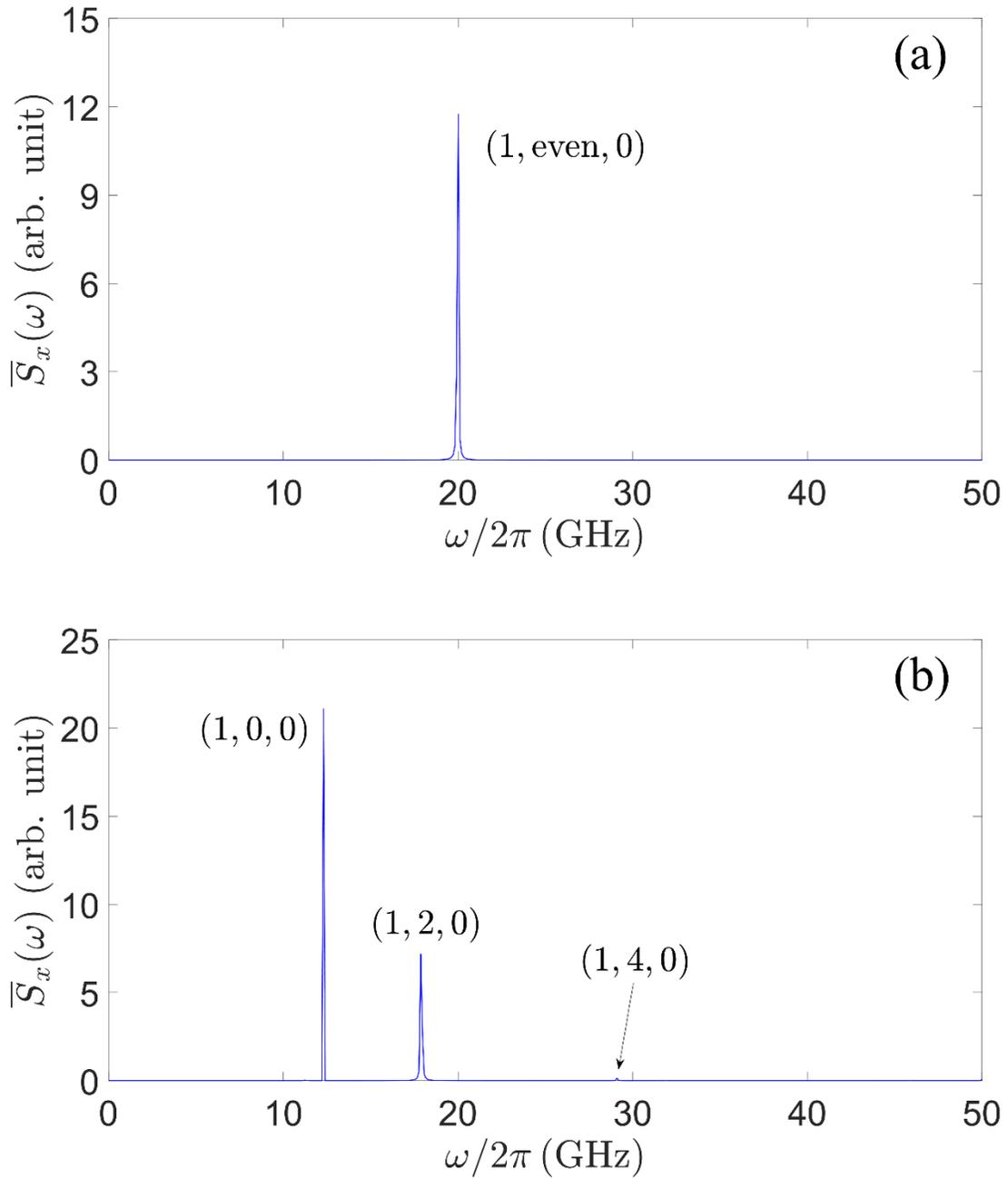

FIG. 9. Power spectral densities of (a) initial nonresonantly driven excitation of a simulation with a microwave field with $p = 1$ symmetry at $f_0 = 20$ GHz; (b) the state that develops from that generated in (a) if the driving field and damping are turned off.

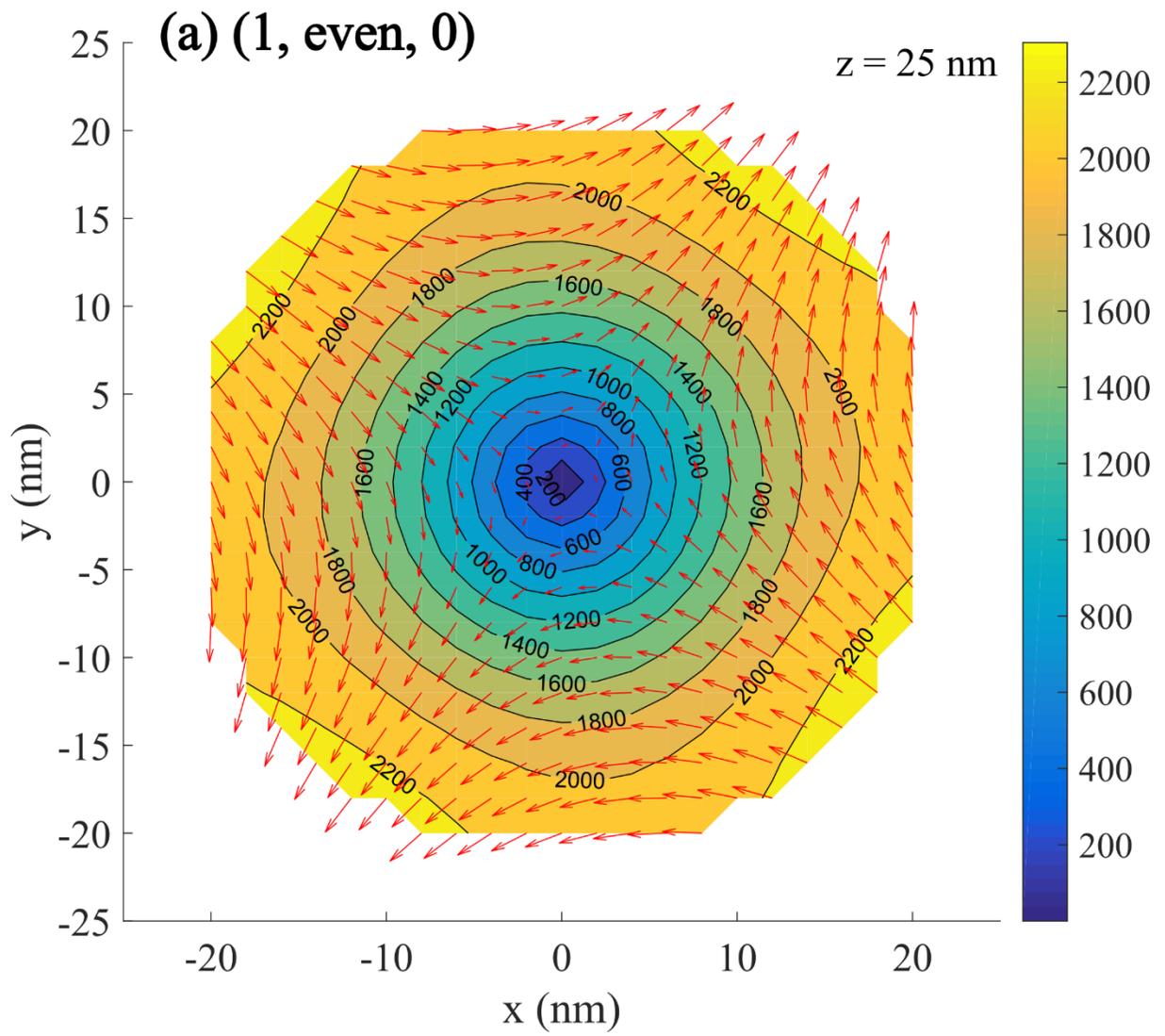

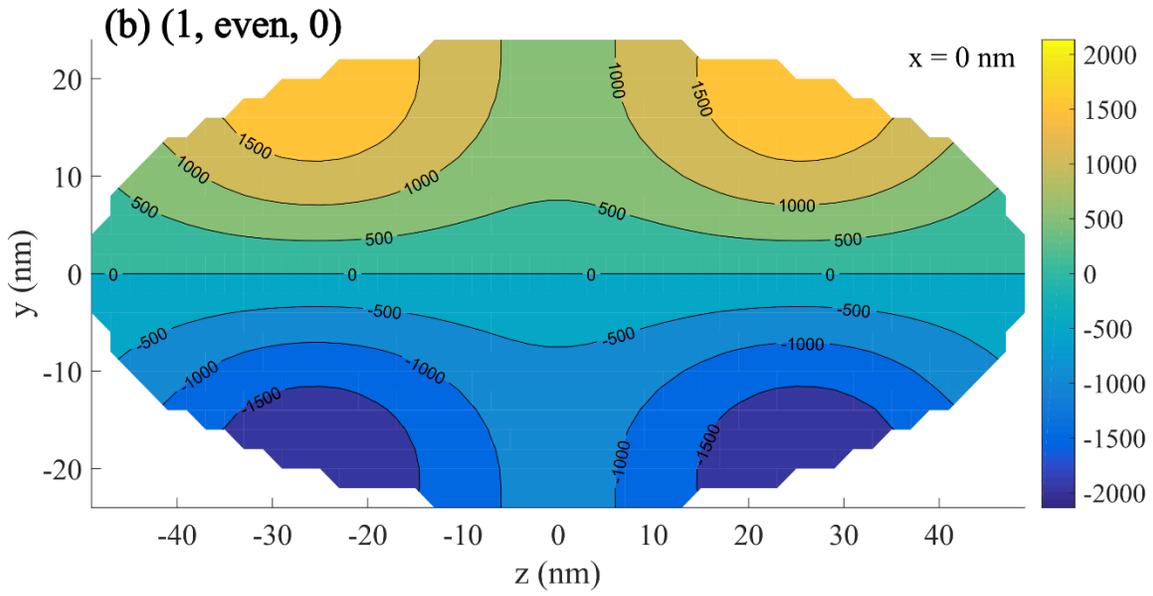

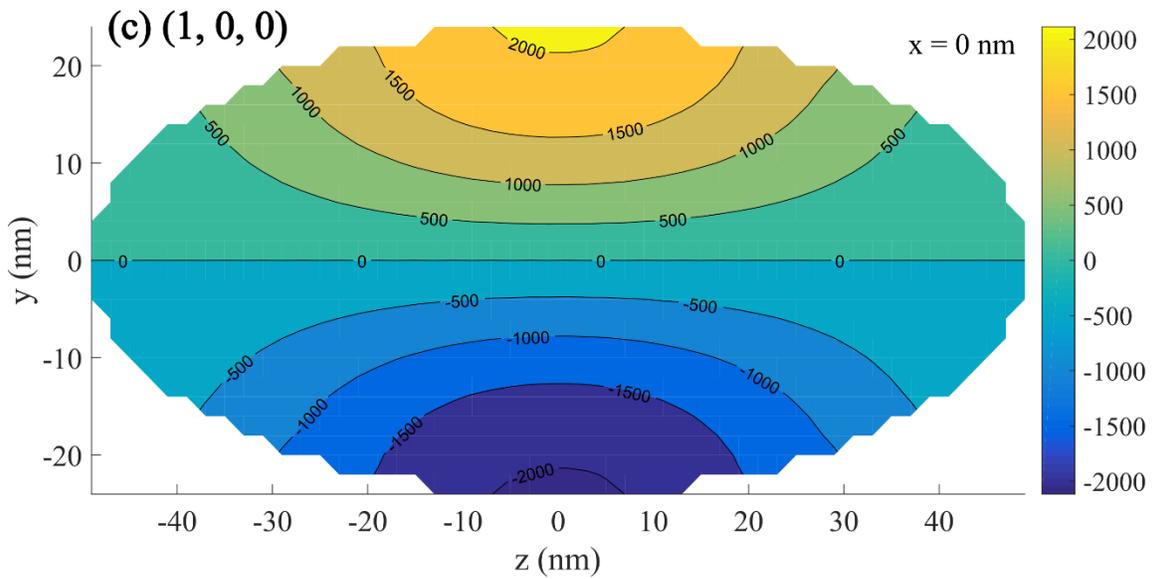

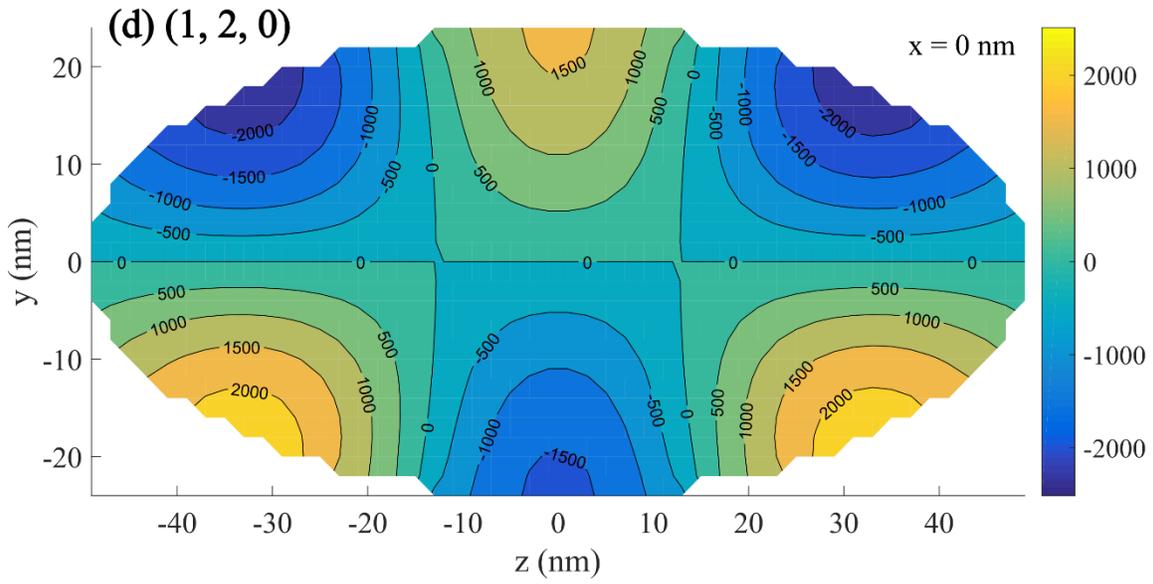

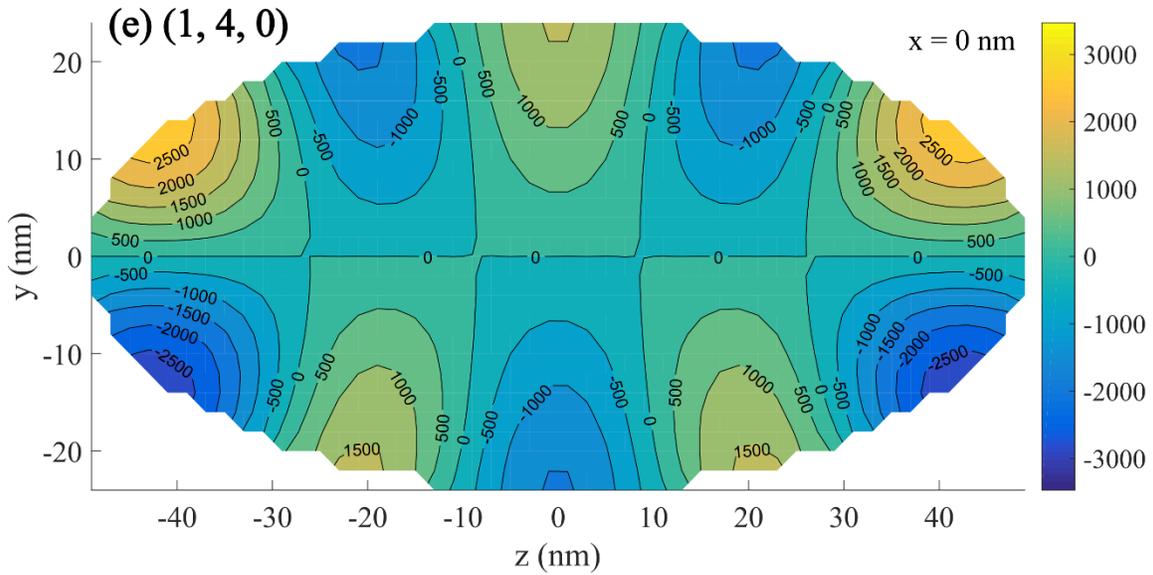

FIG. 10. Associated mode patterns of the simulation for Fig. 9. Figs. 10(a) and (b) shows x-y and y-z cross section of real component of (1, even, 0) mode function respectively; Figs. 10(c), (d), and (e) shows y-z cross section of (1, 0, 0), (1, 2, 0), and (1, 4, 0) mode functions respectively.

The excitation of the sample and analysis of the response follows the same path as in Sec. IV and V. As an example, Fig. 11 shows various plots about an instability corresponding to the Case 3 in Table IV with $f_0 = 20$ GHz and $\dot{H}_1(t) = 13.3$ Oe/ns. Fig. 11(a) shows that $M_x$ doesn't have noticeable changes because all initial and final modes have p = ±1 in which summing over the entire volume yields $M_x = M_y = 0$. By investigating the PSD and associated mode patterns, which are partially shown at Figs. 11(b) and (c), we identified which modes are excited at the instability. The mode patterns of (−1, 0, 0) can be seen at Fig3. (d) and (e).

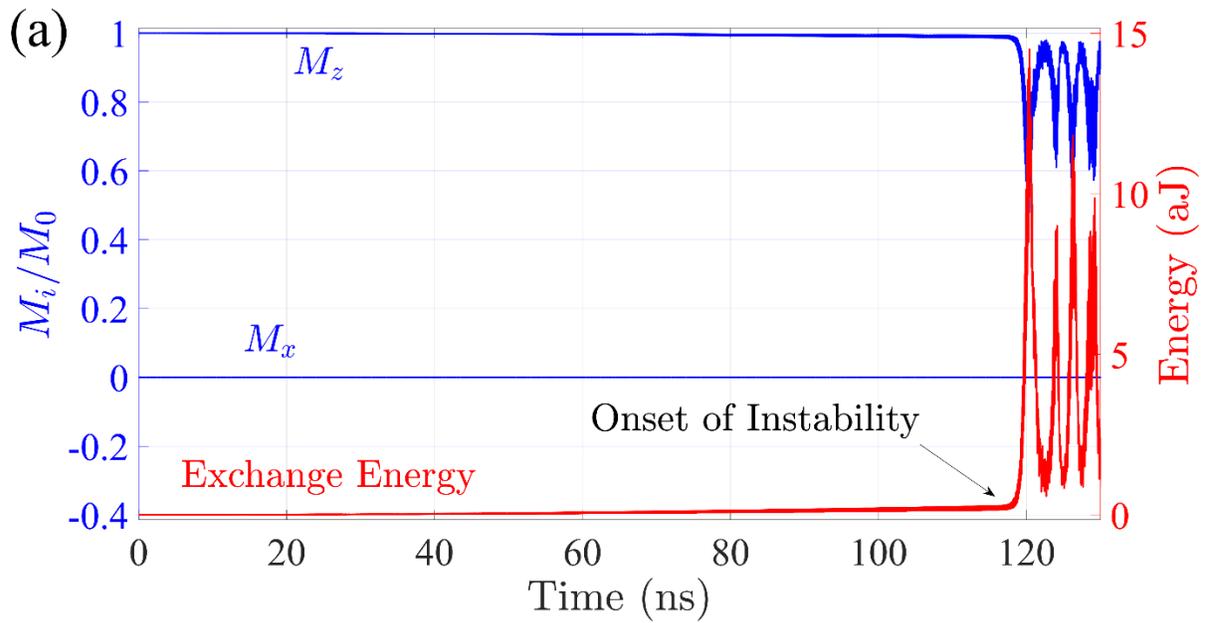

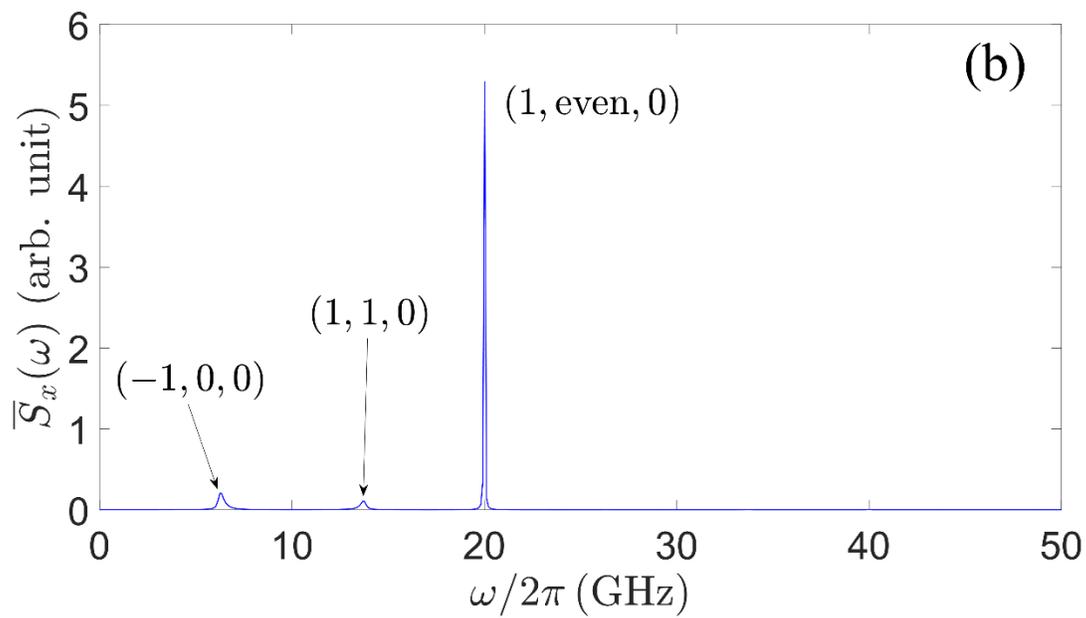

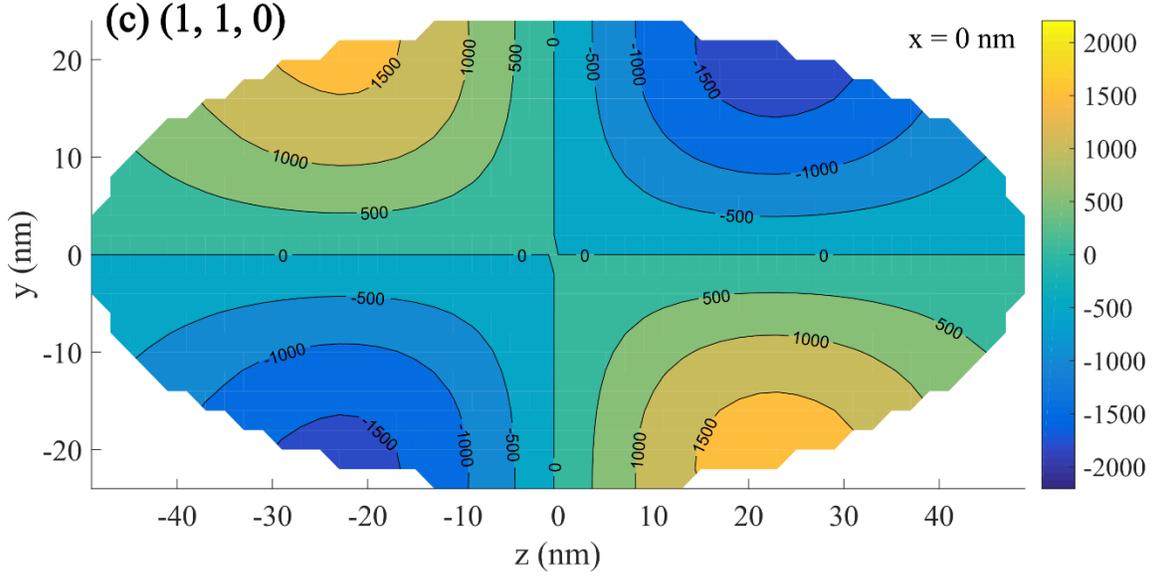

FIG. 11. A number of plots associated with an instability of Case 3 in Table IV at $f_0 = 20$ GHz: (a) raw data which shows an onset of the instability around 117 ns, (b) PSD for 109–119 ns, and (c) y-z cross section of imaginary value of (1, 1, 0) mode amplitude.

Since $p_i = 1$, we expect $p_{f1} + p_{f2} = 0$. There are two main ways in which this can happen: $p_{f1} = 1$ and $p_{f2} = -1$, or $p_{f1} = p_{f2} = 0$. We see both. We also see one process in which $p_{f1} = -p_{f2} = 2$ (Case 18 in Table IV).

Table IV shows three-mode processes we have examined with a $p = 1$ initial state. The format is the same as for Table II and III. One can still find the two trends which we identified for Table III and one can find another trend here which is that a smaller $n_z$ (or $n_{z1} + n_{z2}$) is preferred. For example, one can notice this third trend by looking at cases of $n_{z1}(p_{z1} = 1) - n_{z2}(p_{z2} = -1) = -1$, $n_{z1}(p_{z1} = 0) - n_{z2}(p_{z2} = 0) = -1$, and $n_{z1}(p_{z1} = 0) - n_{z2}(p_{z2} = 0) = -3$ because these cases indicate instabilities are only observed when $n_z$ (or $n_{z1} + n_{z2}$) is lower than some numbers. These trends for thresholds can be roughly understood in terms of the overlap integral of the associated three modes and mediating interactions, because having smaller $|n_{z1} - n_{z2}|$, sharing the same $p$ between initial and final states, and having smaller $n_{z1} + n_{z2}$ all generally help to increase the integral.

The suggested three trends are generally able to distinguish observed and unobserved instabilities, but these don't explain why a case of $n_{z1}(p_{z1} = 0) - n_{z2}(p_{z2} = -1) = -1$ is unobservable while a case of $n_{z1}(p_{z1} = 0) - n_{z2}(p_{z2} = -1) = 1$ is observable in Table II. To resolve this anomaly, we speculate that one needs to conduct explicit calculation of third order dipolar energy, (A.21), responsible for the three-mode instabilities, with given mode functions from the micromagnetic simulation.

Table IV. Observed and unobserved three-mode down conversion processes excited by an initial nonuniformly precessing state which can excite $p = 1$ modes. The first column lists the quantum numbers $(p_1, n_{z1})$ and $(p_2, n_{z2})$ of the two final-state modes; $n_r = 0$ for all modes except Case 19. All processes tabulated satisfy the selection rules (3.1)–(3.3).

| Case No. | Final state modes | | Frequencies (GHz): $f_{observed}$ ($f_{linear}$) | | | Observed? | Comments |
|---|---|---|---|---|---|---|---|
| | $(p_1, n_{z1})$ | $(p_2, n_{z2})$ | $f_0$ | $f_1$ | $f_2$ | | |
| 1 | (1, 0) | (−1, 1) | 22.00 | 11.33 (12.3) | 10.64 (11.23) | Yes | $\dot{H}_1(t) = 40$ Oe/ns $\alpha = 8 \times 10^{-3}$ & $1 \times 10^{-4}$. |
| 2 | (1, 0) | (−1, 3) | 33.11 | (12.3) | (21.88) | No | |
| 3 | (1, 1) | (−1, 0) | 20.00 | 13.67 (14.36) | 6.39 (6.45) | Yes | Also seen at $f_0 = 16.0$ & 16.5 GHz |
| 4 | (1, 1) | (−1, 2) | 29.00 | 13.48 (14.36) | 15.53 (16.21) | Yes | |
| 5 | (1, 2) | (−1, 1) | 27.00 | 16.5 (17.87) | 10.45 (11.23) | Yes | |
| 6 | (1, 2) | (−1, 3) | 39.00 | (17.87) | (21.88) | No | |

| | | | | | | |
|---|---|---|---|---|---|---|
| 7 | (1, 3) | (−1, 0) | 29.00 | (22.95) | (6.45) | No | |
| 8 | (1, 3) | (−1, 2) | 36.50 | 21.39 (22.95) | 15.14 (16.21) | Yes | |
| 9 | (0, 0) | (0, 1) | 8.50 | 4.20 (5.66) | 4.20 (4.98) | Yes | Broad peaks |
| 10 | (0, 0) | (0, 3) | 14.00 | 4.79 (5.66) | 9.18 (10.06) | Yes | |
| 11 | (0, 0) | (0, 5) | 27.50 | (5.66) | (22.17) | No | |
| 12 | (0, 1) | (0, 2) | 11.50 | 6.93 (4.98) | 4.69 (6.64) | Yes | Strange frequency shifts, (0, 2) & (0, 0) mixed at 4.69 |
| 13 | (0, 1) | (0, 4) | 20.00 | (4.98) | (15.33) | No | |
| 14 | (0, 2) | (0, 3) | 16.00 | 6.35 (6.64) | 9.67 (10.06) | Yes | $\alpha = 1\times 10^{-4}$ |
| 15 | (0, 2) | (0, 5) | 27.50 | (6.64) | (22.17) | No | |
| 16 | (0, 3) | (0, 4) | 25.00 | (10.06) | (15.33) | No | |
| 17 | (0, 4) | (0, 5) | 37.20 | (15.33) | (22.17) | No | |
| 18 | (2, 1) | (−2, 0) | 42.00 | 24.02 (25.59) | 17.97 (19.04) | Yes | |
| 19 | (0, 3) | (0, 0, $n_r$ = 1) | 36.50 | 9.57 (10.06) | 26.95 (28.42) | Yes | $\alpha = 1\times 10^{-4}$ |

A. Discussion of individual cases

As for $p = -1$, we only discuss selected cases.

**Case 1**. The final state sought is (1, 0) + (−1, 1), with $f_1 + f_2 = 12.3 + 11.23 = 23.53$ GHz. A run at $f_0 = 23.5$GHz showed no instability, even if damping was set to zero. In this case, OOMMF was run in a mode that explicitly enforces energy conservation after accounting for the energy fed in by the driving field, $\mathbf{H}_{mw}$. However, it was seen at $f_0 = 22$GHz with $\dot{H}_1 = 40$ Oe/ns and $\alpha = 10^{-4}$ and $8\times10^{-3}$ with $f'_{1,2}$ as in Table IV, suggesting a small coupling constant and a correspondingly large threshold.

This state was also seen in a simulation at $f_0 = 23$ GHz with $\dot{H}_1 = 40$ Oe/ns. However, the value of $n_z$ in what should have been the (1,0) mode was somewhat indefinite. Further, the state was seen rather late, at 90 ns (by which time $H_1 = 3600$Oe), along with upconversion to a (3, 3) mode at 46GHz, which subsequently decayed to (3, 1) + (−1, 1) at 35.5 + 10.4 GHz. These two stages were seen via FFT's over different time intervals. The tipping angle (which is obviously inhomogeneous) was as large as 13.5°.

When the same simulation was allowed to run till 130 ns (by which time $H_1 = 5200$Oe), all the above peaks were seen in the FFT. However, the $n_z$ quantum number for the 46GHz peak was not definitely 3; all we can say is that the mode is odd in z. In addition, several other peaks were seen which we group as follows (all frequencies in GHz):

$$\begin{aligned}&\text{A: } (1, 3) \text{ at } 16.9 + (-1, 0) \text{ at } 6.2\\&\text{B: } (1, \text{even}) \text{ at } 10.7 + (-1, 1) \text{ at } 12.2\\&\text{C: } (-1, 0) \text{ at } 6.2 + (3, 2) \text{ at } 39.9\\&\text{D: } (-1, 1) \text{ at } 12.2 + (3, 1) \text{ at } 33.8\end{aligned} \qquad (6.3)$$

Process A is Case 7, with a very large (~6 GHz) downshift. Process B could be a highly noresonantly driven transition to something like Case 5, (1, 2) + (−1, 1). Processes C and D appear to be two-mode decays of the 46 GHz mode. Note that these two processes involve a mode which is also present in one of the two-mode decays of the initial state.

**Case 2**. Here, with $f_0 = 34.18$GHz, we sought and failed to see the final state (1, 0) + (−1, 3), suggesting that processes with large $|n_{z1} - n_{z2}|$ are suppressed. We also ran a simulation with $f_0 = 33.11$ GHz which is the frequency sum for the final state (−1, 1) + (−1, 3) with the goal of testing the selection rules, since this final state violates both the *p* and *z*-parity rules. No instability was seen.

**Case 3**. The final state is (1, 1) + (−1, 0) with $f_1 + f_2 = 14.36 + 6.45 = 20.81$ GHz. A run at $f_0 = 20.81$ GHz showed no instability, even with $\alpha = 3 \times 10^{-4}$ or $10^{-4}$. However, this process is very sensitive to $f_0$, and it was seen at $f_0 = 20.00$ GHz with $f'_{1,2} = 13.67$ and 6.35 GHz. The frequency sum of 20 GHz is very close to that for the state (0, 1) + (0, 4) (Case 13), which we did not see; only Case 3 is seen.

The Case 3 final state was also seen in several other runs: (i) at $f_0 = 16.5$ GHz, with significant frequency downshifts ($f'_{1,2} = 12.3$ and 4.2–4.3 GHz); (ii) at $f_0 = 16.5$ GHz with $\alpha = 10^{-4}$, showing similar downshifts; (iii) at $f_0 = 16.0$ GHz with $f'_{1,2} = 11.43$ and 4.49 GHz. In this connection see also Case 14.

**Cases 4 and 5**. The final states are (1, 1) + (−1, 2) (Case 4) and (1, 2) + (−1, 1) (Case 5), both of which are seen as described in Table IV. This suggests that when $p_{f2} = -p_{f1} = 1$, $n_{z1} - n_{z2}$ can have either sign. Note that the frequency sum for these two cases is quite close.

A frequency sum of 27.46 GHz is obtained for the final state (1, 0) + (−1, 2) which violates the *z*-parity rule. We failed to see this state with $f_0 = 27$ GHz even after adding a (1, 0) seed. Instead, only the Case 5 final state was seen, providing a test of the selection rule. Similarly, a run at 27.5 GHz is potentially capable of showing an instability to (0, 0) + (0, 5) (Case 11). We did not see this and only saw Case 5 instead.

The final state for Case 4 is nearly degenerate with (1, 3) + (−1, 0) (Case 7), with $f_1 + f_2$ being 30.57 GHz for Case 4 and 29.4 GHz for Case 7. We failed to see the Case 7 state even after seeding with a (−1, 0) mode with $f_0 = 29$ GHz, again suggesting that processes with large $|n_{z1} - n_{z2}|$ are suppressed. Instead, only Case 4 was seen.

**Case 7**. The final state is (1, 3) + (−1, 0) with $f_{1,2}$ = 22.95 and 6.45 GHz. We failed to see this state in a direct simulation at $f_0$ = 29.0 GHz, and also in a simulation that seeded the (−1, 0) mode as discussed in connection with Case 4. However, we did see a state with modes having these quantum numbers in the very highly driven simulation at 23GHz discussed in connection with Case 1.

**Case 8**. The final state is (1, 3) + (−1, 2) with a linear regime frequency sum of 39.2 GHz. A run at 39 GHz with $\dot{H}_1$ = 40 Oe/ns showed no instability. A second run at 36.5 GHz did show this final state; the nearby state (0, 4) + (0, 5) — Case 17 — was not seen. In a third run at 36.50 GHz and $\alpha = 10^{-4}$, this instability was preempted by one to the pair

$$(0, 3) \text{ at } 9.57 \text{ GHz} + (0, 0, n_r = 1) \text{ at } 26.95 \text{GHz}, \tag{6.4}$$

with only modest frequency downshifts. This is the only Case where we see a final state to a mode with $n_r \neq 0$, and it is listed as Case 19 in the Table IV.

**Case 9**. Here, our goal was to see the final state (0, 0) + (0, 1) with $f_1 + f_2 = 5.66 + 4.98 = 10.64$ GHz. A run at $f_0 = 10.5$ GHz showed an instability but the FFT and spin maps of the resulting state did not reveal clear modes. A second run at $f_0 = 8.5$ GHz and $\dot{H}_1 = 40$ Oe/ns showed a very broad peak centered at 4.20GHz, and spin maps indicative of the (0, 0) and (0, 1) modes. We suspect that the near degeneracy of the final state modes is giving rise to irregular transfer of energy between them and broadening the associated frequency, and that we do not get a well-defined two-mode state before the system goes chaotic.

**Case 10**. We chose $f_0 = 14$ GHz and saw the intended final state (0, 0) + (0, 3) with frequencies as listed in Table IV. There is a modest pulling down of frequencies (about 0.9 GHz). We also saw this case in a run at 15GHz, where we were looking for, and did not find, a decay to the parity-selection-rule-violating final state (0, 1) + (0, 3). Note that the final state (0, 1) + (0, 3) also can be test by $f_0$ = 14 GHz where we only saw Case 10.

**Case 12**. We chose $f_0$ = 11.50 GHz and saw the intended final state (0, 1) + (0, 2), but the frequency shifts were peculiar. The states (0, 1) and (0, 2) were significantly upshifted and downshifted respectively. The mode pattern at $f'_2$ = 4.69 GHz is a mixed pattern between (0, 0) and (0, 2) with about 1:1 amplitude.

One notable phenomenon, occurring at $f_0$ = 10.50, 11.00, and 11.50 GHz, is that the nonresonantly driven peak at $f_0$ is broadened or has satellites before the three-mode instability happens. The mode (1, 0) has a linear regime frequency of 12.3 GHz. We speculate that since the natural frequency of (1, 0) generally decreases as the excitation level increases, $\mathbf{H}_{mw}$ is resonantly exciting this mode for some period of time. As $H_1(t)$ increases further, we have $f_{(1,0)}(t) < f_0$, and the resonantly excited mode (1, 0) becomes a transient mode, but it doesn't decay immediately. That is why the observed broadening mainly occurs at the lower frequency side of $f_0$.

The mixed mode pattern for the state nominally identified as (0, 2) also appears to be connected to Case 9. Note that $f_{(0,0)} + f_{(0,1)}$ = 10.64 GHz which is close to $f_{(0,1)} + f_{(0,2)}$ = 11.62 GHz.

**Case 14**. Here the final state is (0, 2) + (0, 3), with $f_1 + f_2$ = 6.64 + 10.06 = 16.7 GHz. The Case 3 simulation at $f_0$ = 16 GHz was rerun with $\alpha = 10^{-4}$, and the Case 3 final state, (1, 1) + (−1, 0) was preempted by an instability to Case 14 with $f'_{1,2}$ = 6.35 and 9.67 GHz.

**Case 16**. This final state, (0, 3) + (0, 4), was not seen in a run at $f_0$ = 25 GHz with $\dot{H}_1$ = 13.3 Oe/ns. Increasing $\dot{H}_1$ to 40 Oe/ns showed upconversion to the (3,3) mode at 50GHz at a time of about 90 ns, followed by the following two-mode decays at a very late time (125 ns):

A: (−1, 1) at 9.8 GHz + (3, 1) at 40.2 GHz

$$\text{B: }(1, 2) \text{ at } 15.2 \text{ GHz} + (1, \text{even}) \text{ at } 34.9 \text{ GHz} \tag{6.5}$$

The same upconversion was also seen in the simulation discussed for Case 1 ($f_0 = 22.5$ and 23 GHz and $\dot{H}_1 = 40$ Oe/ns).

**Case 18**. This is a very clean process with details as shown in Table IV. It is noteworthy because $p_{f1,2} = \pm 2$.

## VII. Conclusions

We have conducted a comprehensive study of Suhl instabilities in a nano-scale spheroidal ferromagnet and provided generalized symmetry-related selection rules with proper quantum numbers. One of the results was that OAM was not conserved during 3-magnon processes. Among allowed instabilities, we have identified general trends, which can be understood from the overlap integral that distinguishes between favored (lower threshold) and unfavored (higher threshold) processes. However, we also found anomalies that don't fit the general trends. These could be studied further by numerically calculating coupling constants and mode-dependent damping. In particular, mode-dependent dampings might be another factor to explain the unobserved cases.

The Suhl instability is one of the first nonlinear phenomena encountered when going from the linear to the nonlinear regime in nano-scale confined geometries. In many cases, there is enough parameter space to carry out systematic studies before the onset of other highly nonlinear phenomena (e.g., chaotic response, vortex nucleation, etc.). To study this shallow nonlinear regime, one needs to characterize Suhl instabilities, especially the three-mode instability which, in general, occurs first. Since this instability is solely mediated by the dipolar interaction, this interaction is an essential component of the study of nonlinear phenomena in magnetic systems with nanoscale confined geometries.

As nanofabrication techniques have been developed and advanced, much research has been carried out in magnetic nano-scale bounded systems (or nanoparticles), including 3D nanomagnetism [51-57], magnetic nanodots [81-84], artificial spin ice [85-91], magnonic crystals [92-97], magnonic devices [98-101], and microwave assisted magnetization reversals [102-114]. Our study can be regarded as an initial attempt to connect the linear regime and highly-nonlinear regime of magnetic systems with nano-scale 3D confined geometries.

Acknowledgement

This research was carried out under the support of U.S. Department of Energy through grant DE-SC0014424. This research was supported in part through the computational resources and staff contributions provided for the Quest high performance computing facility at Northwestern University which is jointly supported by the Office of the Provost, the Office for Research, and Northwestern University Information Technology.

APPENDIX A.    The selection rules for three-mode instability

In this Appendix, we derive the selection rules for three-mode-processes. Suhl gave these rules without explanation when the initial mode is the uniform FMR mode ($p = n_z = n_r = 0$). Thus, this Appendix may be useful both for readers who wish to see an explanation, and for the generalization to arbitrary initial states. Although our analysis is done using the normal modes in the linear regime (small tipping angles), it is clear that the selection rules themselves depend only on the symmetries of the modes and should therefore extend to the moderately nonlinear regime also.

The selection rules depend critically on the structure of the third order term when the total energy is expanded in terms of the normal modes, so we examine this in the next subsection.

A. General expression for total energy

The energy consists of three pieces:

$$\mathcal{E} = \mathcal{E}_Z + \mathcal{E}_{\text{ex}} + \mathcal{E}_{\text{dd}}, \tag{A.1}$$

the Zeeman energy, the exchange energy, and the dipole-dipole interaction. The Zeeman energy with an applied field $\mathbf{H}_0 \parallel \hat{z}$ is

$$\mathcal{E}_Z = -\int d^3 r \mathbf{H}_0 \cdot \mathbf{M}. \tag{A.2}$$

We are writing $\mathbf{M}(\mathbf{r})$ for the magnetization, with $|\mathbf{M}(\mathbf{r})| = M_0$ at every $\mathbf{r}$. The exchange energy is

$$\mathcal{E}_{\text{ex}} = \frac{A_{\text{ex}}}{M_0^2} \int d^3 r (\nabla_i M_j)(\nabla_i M_j). \tag{A.3}$$

The dipole-dipole energy is given by

$$\mathcal{E}_{\text{dd}} = \frac{1}{2} \iint d^3 r \, d^3 r' M_i(\mathbf{r}) K_{ij}(\mathbf{r} - \mathbf{r}') M_j(\mathbf{r}'), \tag{A.4}$$

where

$$K_{ij}(\mathbf{R}) = \frac{R^2 \delta_{ij} - 3 R_i R_j}{R^5} + \frac{4\pi}{3} \delta_{ij} \delta(\mathbf{R}). \tag{A.5}$$

The delta function term is necessary in a continuum formulation of the magnetization. Alternatively, we may think in terms of the pseudoelectrostatic energy of interaction between volume and surface magnetic charges $-\nabla \cdot \mathbf{m}$ and $\mathbf{m} \cdot \hat{\mathbf{n}}$. This leads to

$$\mathcal{E}_{\text{dd}} = \frac{1}{2} \int_{V^+} \int_{V^+} d^3 r \, d^3 r' \left( -\nabla \cdot \mathbf{M}(\mathbf{r}) \right) \frac{1}{|\mathbf{r} - \mathbf{r}'|} \left( -\nabla' \cdot \mathbf{M}(\mathbf{r}') \right). \tag{A.6}$$

Here $\nabla \cdot \mathbf{M}$ is a generalized divergence implemented by extending the volume of integration slightly beyond the sample, as indicated by the notation $V^+$. The surface poles are thereby included. Yet another way of writing this expression is

$$\varepsilon_{dd} = \frac{1}{2} \int\int d^3r\, d^3r' M_i(\mathbf{r}) \bar{\partial}_i \frac{1}{|\mathbf{r}-\mathbf{r}'|} \vec{\partial}_j M_j(\mathbf{r}'). \tag{A.7}$$

In this form the dipole-dipole interaction kernel is

$$K_{ij}(\mathbf{R}) = \bar{\partial}_i \frac{1}{|\mathbf{R}|} \vec{\partial}_j. \tag{A.8}$$

Viewed as an operator, $K$ is Hermitian in the sense that if we interpret

$$(f, K_{ij}g) = \int\int d^3r\, d^3r' f^*(\mathbf{r}) K_{ij}(\mathbf{r}-\mathbf{r}') g(\mathbf{r}'), \tag{A.9}$$

then

$$(f, K_{ij}g)^* = (g, K_{ij}f). \tag{A.10}$$

In short, $K_{ij}^\dagger = K_{ji}$.

1. The demagnetization field

In the compact form $(f, K_{ij}g)$ we can replace $f$ and $g$ by vector fields, $\mathbf{f}(\mathbf{r})$ and $\mathbf{g}(\mathbf{r})$, which we can think of as magnetization configurations. Then from the energy expression, we see that we can write (implicit sum on $i$ and $j$ indices)

$$(f_i, K_{ij}g_j) = -\mathbf{f} \cdot \mathbf{h}_d[\mathbf{g}], \tag{A.11}$$

where $\mathbf{h}_d[\mathbf{g}]$ is the demagnetization field generated by $\mathbf{g}$:

$$h_{d,i}[\mathbf{g}] = -\int d^3r' K_{ij}(\mathbf{r}-\mathbf{r}') g_j(\mathbf{r}'). \tag{A.12}$$

In particular, if **g** is uniform, for an ellipsoid we get

$$(f_i, K_{ij} 1) = 4\pi f_i N_{ij}, \qquad (A.13)$$

where $N_{ij}$ is the demagnetization tensor.

### B. Small tipping angles

When the tipping angle is small, we write **m** for the $xy$ plane part of $\mathbf{M}(\mathbf{r})/M_0$, and

$$\mathbf{u} = \left( \mathbf{m} + \left(1 - m^2\right)^{1/2} \hat{\mathbf{z}} \right), \qquad (A.14)$$

Thus,

$$u_x = m_x, \quad u_y = m_y, \quad u_z = 1 - \frac{1}{2}m^2 - \frac{1}{8}\left(m^2\right)^2 + \cdots. \qquad (A.15)$$

To zeroth order we put $\mathbf{m} = 0$, i.e., $\mathbf{M}(\mathbf{r}) = M_0 \hat{\mathbf{z}}$. Then the exchange energy is zero, and the dipole-dipole energy becomes

$$\mathcal{E}_{dd}^0 = \frac{1}{2}(M_0, K_{zz} M_0) = 2\pi V N_{zz} M_0^2. \qquad (A.16)$$

The Zeeman energy is $\mathcal{E}_Z^0 = -M_0 H_0 V$.

There is no term in the energy linear in $m_x$ or $m_y$. The quadratic terms in the Zeeman and exchange energies are

$$\mathcal{E}_Z^{(2)} = \frac{1}{2} M_0 H_0 \int d^3r \left(m_x^2 + m_y^2\right), \qquad (A.17)$$

$$\mathcal{E}_{ex}^{(2)} = A_{ex} \int d^3r \left[ (\nabla m_x)^2 + (\nabla m_y)^2 \right]. \qquad (A.18)$$

For the dipole-dipole energy, letting $u_z'$ be the nonuniform part of $u_z$, we have,

$$\begin{aligned}\mathcal{E}_{dd}^{(2)} &= \frac{1}{2}M_0^2\left[(m_x,K_{xx}m_x)+2(m_x,K_{xy}m_y)+(m_y,K_{yy}m_y)\right]+\frac{1}{2}\times 2(M_0 u'_z, K_{zz}M_0)\\ &= \frac{1}{2}M_0^2\left[(m_x,K_{xx}m_x)+2(m_x,K_{xy}m_y)+(m_y,K_{yy}m_y)\right]\\ &\quad - 2\pi N_{zz}M_0^2 \int d^3r\left(m_x^2+m_y^2\right).\end{aligned} \qquad (A.19)$$

The last term has the same form as the quadratic part of the Zeeman energy, so it may be combined with it by replacing $H_0$ by $H_{\text{in}} = H_0 - 4\pi N_{zz}M_0$.

In the third order, only the dipole-dipole energy contributes and is

$$\mathcal{E}_{dd}^{(3)} = \frac{1}{2}M_0^2 \times 2\left[(m_x, K_{xz}u'_z) + (m_y, K_{yz}u'_z)\right]. \qquad (A.20)$$

An expansion to higher order, done in terms of the Landau-Lifshitz-Gilbert equation rather than the total energy, can be found in Ref. [79].

### C. Selection rules for three-mode processes

Written out in more detail, the third order term in the energy/Hamiltonian is

$$\mathcal{E}^{(3)} = -\frac{1}{2}M_0^2 \iint d^3r\, d^3r'\left[m_x(\mathbf{r})K_{xz}(\mathbf{r}-\mathbf{r}') + m_y(\mathbf{r})K_{yz}(\mathbf{r}-\mathbf{r}')\right]m_\perp^2(\mathbf{r}'). \qquad (A.21)$$

In an ellipsoidal sample, the energy is invariant under z-axis rotations, so total angular momentum along z is conserved. For reflections, there are two operations that one can think about:

$$\text{Full parity:} \quad \mathbf{r},\mathbf{r}' \to -\mathbf{r},-\mathbf{r}'; \qquad (A.22)$$

$$xy \text{ plane reflection:} \quad z, z' \to -z, -z'. \qquad (A.23)$$

(Full parity is the same as a 180° rotoreflection, but it provides insight to examine both operations, even though the conclusions must agree.) Now,

$$K_{xz}(\mathbf{r}-\mathbf{r}') = -3\frac{(x-x')(z-z')}{|\mathbf{r}-\mathbf{r}'|^5},\quad K_{yz}(\mathbf{r}-\mathbf{r}') = -3\frac{(y-y')(z-z')}{|\mathbf{r}-\mathbf{r}'|^5}. \qquad (A.24)$$

The energy is invariant under full parity and changes sign under z-parity (*xy* plane reflection).

Suppose three modes α, β and γ are involved. We can write the integrand as (implicit sum only on $j = x, y$)

$$\left(m_j^{(\alpha)}, K_{jz}\mathbf{m}^{(\beta)} \cdot \mathbf{m}^{(\gamma)}\right) \quad \text{or} \quad \left(m_j^{(\beta)}, K_{jz}\mathbf{m}^{(\alpha)} \cdot \mathbf{m}^{(\gamma)}\right). \tag{A.25}$$

We take the mode functions in the exchange dominated limit as in Lim, Ketterson, and Garg [63]:

$$\begin{pmatrix} m_x \\ m_y \end{pmatrix} = \frac{1}{\sqrt{2}} \begin{pmatrix} 1 \\ i \end{pmatrix} F(r,z) e^{ip\varphi} \times e^{-i\omega t}, \tag{A.26}$$

where $\omega > 0$, and $r$, $z$, $\varphi$ are cylindrical coordinates. This has angular momentum $m_{\text{orbital}} = p$, $m_{\text{spin}} = 1$, and $m_{\text{total}} = p+1$. Let us consider the process

$$\alpha \to \beta + \gamma. \tag{A.27}$$

Then, we obviously need

$$\omega_\alpha = \omega_\beta + \omega_\gamma. \tag{A.28}$$

Further, since the total angular momentum is $p + 1$, we need

$$p_\alpha + 1 = (p_\beta + 1) + (p_\gamma + 1), \tag{A.29}$$

i.e.,

$$p_\beta + p_\gamma = p_\alpha - 1. \tag{A.30}$$

For parity let us consider full and z-parity separately. Suppose

$$F(r,-z) = (-1)^\tau F(r,z), \tag{A.31}$$

where τ is either 0 or 1, and is the z-parity of a mode. Under full parity, $\varphi \to \varphi + \pi$, so

$$e^{ip\varphi} \to (-1)^p e^{ip\varphi}, \tag{A.32}$$

and the full parity of a mode is $\tau + p$ (mod 2).

So under full parity, we need (all sums mod 2)

$$\left(\tau_\beta + p_\beta\right) + \left(\tau_\gamma + p_\gamma\right) = \tau_\alpha + p_\alpha. \tag{A.33}$$

Under z-parity on the other hand, because $K_{xz}$ and $K_{yz}$ are odd, we want

$$\text{i.e., } \tau_\beta + \tau_\gamma = \tau_\alpha + 1. \tag{A.34}$$

If we combine the full parity rule with the angular momentum rule, we see these are the same, and we can state the rule entirely in terms of reflection in $z$:

$$\alpha = \text{even:} \qquad (\beta, \gamma) = \text{(one even, one odd)}, \tag{A.35}$$

$$\alpha = \text{odd:} \qquad (\beta, \gamma) = \text{(both even, or both odd)}. \tag{A.36}$$

In particular, if $\alpha$ is the uniform mode, and we consider its decay to $\beta + \gamma$, the relevant piece of the cubic energy is

$$\left(m_j^{(\beta)}, K_{jz}\mathbf{m}^{(unif)} \cdot \mathbf{m}^{(\gamma)}\right). \tag{A.37}$$

Taking $F^{(unif)} = 1$, $p^{(unif)} = 0$, this piece equals

$$\frac{M_0^2}{2 \times 2^{3/2}} \int_{\mathbf{r}} \int_{\mathbf{r}'} \left(K_{xz} - iK_{yz}\right) F_\beta(r,z) \times 2F_\gamma(r',z')$$

$$= \frac{M_0^2}{2^{3/2}} \int (-3) \frac{\left(re^{-i\varphi} - r'e^{-i\varphi'}\right)(z-z')}{\left[\left(r^2 - 2rr'\cos(\varphi - \varphi') + r'^2\right) + (z-z')^2\right]^{5/2}} F_\beta(r,z) F_\gamma(r',z') e^{-i\left(p_\beta \varphi + p_\gamma \varphi'\right)}$$

$$\times rr'\, dr\, dr'\, dz\, dz'\, d\varphi\, d\varphi'. \tag{A.38}$$

If we put $\varphi = \varphi' + \alpha$, the $\varphi'$, $\alpha$ part of the integrand is

$$\frac{\left(re^{-i\alpha} - r'\right)e^{-i\varphi'}}{\left[\left(r^2 - 2rr'\cos\alpha + r'^2\right) + (z-z')^2\right]^{5/2}} e^{-i\left(p_\beta + p_\gamma\right)\varphi'} e^{-ip_\beta \alpha}. \tag{A.39}$$

When we integrate this over $\varphi'$, we get zero unless $p_\beta + p_\gamma = -1$, exactly the selection rule we found before. The remaining integral can be written using this rule as

$$-\frac{3M_0^2}{2^{3/2}} \int \frac{\left(re^{-ip_\gamma\alpha} - r'e^{-ip_\beta\alpha}\right)(z-z')}{\left[\left(r^2 - 2rr'\cos\alpha + r'^2\right) + (z-z')^2\right]^{5/2}} F_\beta(r,z) F_\gamma(r',z') rr' \, dr \, dr' \, dz \, dz' \, d\alpha. \quad (A.40)$$

We can also see the z-parity rule from this: the integral vanishes unless modes β and γ have opposite parity in z. Integrals such as this give us the coupling constants in a quantum mechanical formulation wherein the relevant term in the Hamiltonian is written as

$$\sum_{\alpha,\beta,\gamma} \left(g_{\alpha,\beta\gamma} a_\alpha a_\beta^\dagger a_\gamma^\dagger + \text{h.c.}\right), \quad (A.41)$$

where $a_\alpha$ and $a_\alpha^\dagger$ etc. are properly normalized mode annihilation and creation operators.

APPENDIX B.   Model for three-mode instability

We suppose that we have three modes labeled 1, 2, and 3, and consider the process $1 \rightarrow 2 + 3$. Mode 1 is driven by an applied microwave field $\mathbf{H}_{\text{mw}}(t)$ at a frequency $\omega_0$. The equations of motion for the complex amplitudes $a_i$ ($i = 1, 2, 3$) of the modes may then be taken as

$$\dot{a}_1 = -(i\omega_1 + \eta_1)a_1 - iga_2a_3 + \tilde{H}_1 e^{-i\omega_0 t},$$

$$\dot{a}_2 = -(i\omega_2 + \eta_2)a_2 - iga_1 a_3^*, \quad (B.1)$$

$$\dot{a}_3 = -(i\omega_3 + \eta_3)a_3 - iga_1 a_2^*.$$

Here, $\omega_i$ and $\eta_i$ are the frequencies and damping rates of the modes, g is the coupling constant $g_{1,23}$ of Appendix A, and $\tilde{H}_1 \propto \gamma H_1$.

Prior to the instability, we may take $a_{2,3} \ll a_1$, and in steady state, $a_1 = A_1 e^{-i\omega_0 t}$, with

$$|A_1|^2 = \tilde{H}_1^2 / \left[ (\omega_0 - \omega_1)^2 + \eta_1^2 \right] \tag{B.2}$$

We now feed this steady state amplitude in the equations of motion for $a_2$ and $a_3$ and transform to an interaction picture by writing

$$a_{2,3} = c_{2,3} e^{-i\bar{\omega}_{2,3} t}, \tag{B.3}$$

where

$$\bar{\omega}_{2,3} = \omega_{2,3} + \frac{1}{2}\zeta,$$

$$\zeta = \omega_0 - \omega_2 - \omega_3.$$

This leads to equations of motion for $c_2$ and $c_3$, which we write as

$$\begin{aligned}\dot{c}_2 &= \left( i\frac{\zeta}{2} - \eta_2 \right) c_2 - ig A_1 c_3^*, \\ \dot{c}_3^* &= \left( -i\frac{\zeta}{2} - \eta_3 \right) c_3^* + ig A_1^* c_2. \end{aligned} \tag{B.4}$$

Elimination of $c_3^*$ now yields

$$\ddot{c}_2 + (\eta_2' + \eta_3') \dot{c}_2 - \left( g^2 |A_1|^2 - \eta_2' \eta_3' \right) c_2 = 0, \tag{B.5}$$

where $\eta_{2,3}' = \eta_{2,3} \mp i\zeta/2$.

Eq. (B.5) has a solution of the form $e^{st}$ with

$$s = -\frac{1}{2}(\eta_2' + \eta_3') \pm (p + iq), \tag{B.6}$$

with

$$p = \left[ \frac{1}{2} \left( \sqrt{u^2 + v^2} + u \right) \right]^{1/2},$$

$$q = \left[\frac{1}{2}\left(\sqrt{u^2 + v^2} - u\right)\right]^{1/2} \operatorname{sgn}(v),$$

where

$$u = g^2 |A_1|^2 + \frac{1}{4}(\eta_2 - \eta_3)^2 - \frac{1}{4}\zeta^2,$$

$$v = -\frac{1}{2}(\eta_2 - \eta_3)\zeta.$$

An instability will arise if there is a solution with Re($s$) > 0, i.e. $p > (\eta'_2 + \eta'_3)/2$. This condition can be transformed into a threshold for $H_1$, which is

$$H_{1,t} \sim \frac{1}{|g|}\left[(\omega_0 - \omega_1)^2 + \eta_1^2\right]^{1/2}\left[\zeta^2 + (\eta_2 + \eta_3)^2\right]^{1/2} \frac{(\eta_2 \eta_3)^{1/2}}{\eta_2 + \eta_3}. \qquad (B.7)$$

This generalizes Eqs. (25) and (26) of Suhl [4].

We further note that precisely at threshold,

$$q = -\frac{1}{2}\zeta\frac{(\eta_2 - \eta_3)}{(\eta_2 + \eta_3)}. \qquad (B.8)$$

Then the modes 2 and 3 are excited at frequencies $\omega'_{2,3}$ given by

$$\omega'_2 = \omega_2 + \frac{1}{2}\zeta - q,$$
$$\omega'_3 = \omega_3 + \frac{1}{2}\zeta + q, \qquad (B.9)$$

which depart from their natural values $\omega_2$ and $\omega_3$. These shifts depend on the detuning $\zeta$, but also on the damping rates $\eta_2$ and $\eta_3$. Note however that the Manley-Rowe condition is always obeyed:

$$\omega'_2 + \omega'_3 = \omega_2 + \omega_3 + \zeta = \omega_0. \qquad (B.10)$$